\documentclass[a4paper,12pt]{article}

\usepackage[margin=2.5cm]{geometry}

\usepackage{textgreek}

\usepackage[all]{nowidow}
\usepackage{amsmath}
\usepackage[version=4]{mhchem}
\usepackage{siunitx}
\usepackage{float}
\usepackage{graphicx}
\usepackage{caption}
\usepackage{subcaption}
\usepackage[
backend=biber,
style=vancouver
]{biblatex}

\renewbibmacro*{volume+number+eid}{%
    \printfield{volume}%
    \printfield{number}%
}\DeclareFieldFormat[article]{number}{\mkbibparens{#1}}

\DeclareFieldFormat*{pages}{\addspace\mkcomprange{#1}}

\DeclareFieldFormat[article,periodical]{volume}{\mkbibbold{#1}}

\DeclareSIUnit \nucleon {u}
\DeclareSIUnit \ion {ion}
\DeclareSIUnit \frame {frame}
\DeclareSIUnit \Hit {Hit}
\DeclareSIUnit \vertices {vertices}
\DeclareSIUnit \vertex {vertex}

\usepackage{hyperref} 

\graphicspath{{.}}

\newcommand{\cion}{
    \ce{^12C^{6+}}
}

\newcommand{\singleFig}[5][!htbp]{%
    \begin{figure}[#1]%
        \centering%
        \captionsetup{width=15cm}%
        \includegraphics[width=15cm]{#2}%
        \caption[#5]{#4}%
        \label{fig:#3}%
    \end{figure}%
}

\newcommand{\singleFigNarrow}[5][!htbp]{%
    \begin{figure}[#1]%
        \centering%
        \captionsetup{width=15cm}%
        \includegraphics[width=7.9cm]{#2}%
        \caption[#5]{#4}%
        \label{fig:#3}%
    \end{figure}%
}

\newcommand{\doubleFig}[6][!htbp]{%
    \begin{figure}[#1]%
        \centering%
        \captionsetup{width=16cm}%
        \begin{subfigure}[t]{7.9cm}%
            \centering%
            \includegraphics[width=\linewidth]{#2}%
            \phantomsubcaption%
            \label{fig:#4_a}%
        \end{subfigure}%
        \hspace{0.2cm}%
        \begin{subfigure}[t]{7.9cm}%
            \centering%
            \includegraphics[width=\linewidth]{#3}%
            \phantomsubcaption%
            \label{fig:#4_b}%
        \end{subfigure}%
        \caption[#6]{#5}%
        \label{fig:#4}%
    \end{figure}%
}

\newcommand{\quadFig}[8][!htbp]{%
    \begin{figure}[#1]%
        \centering%
        \captionsetup{width=16cm}%
        \begin{subfigure}[t]{7.9cm}%
            \centering%
            \includegraphics[width=\linewidth]{#2}%
            \phantomsubcaption%
            \label{fig:#6_a}%
        \end{subfigure}%
        \hspace{0.2cm}%
        \begin{subfigure}[t]{7.9cm}%
            \centering%
            \includegraphics[width=\linewidth]{#3}%
            \phantomsubcaption%
            \label{fig:#6_b}%
        \end{subfigure}
        
        \begin{subfigure}[t]{7.9cm}%
            \centering%
            \includegraphics[width=\linewidth]{#4}%
            \phantomsubcaption%
            \label{fig:#6_c}%
        \end{subfigure}%
        \hspace{0.2cm}%
        \begin{subfigure}[t]{7.9cm}%
            \centering%
            \includegraphics[width=\linewidth]{#5}%
            \phantomsubcaption%
            \label{fig:#6_d}%
        \end{subfigure}%
        \caption[#8]{#7}%
        \label{fig:#6}%
    \end{figure}%
}

\newcommand{\figref}[1]{
    Figure \ref{fig:#1}
}

\newcommand{\secref}[1]{
    Section \ref{#1}
}

\begin{document}
    \begin{center}
        \Large{Evaluation of a large-area double-sided silicon strip detector for quality assurance in ion-beam radiotherapy}
    \end{center}
    \begin{center}
        \textbf{Devin Hymers}\textsuperscript{1},
        \textbf{Sebastian Schroeder}\textsuperscript{1},
        \textbf{Olga Bertini}\textsuperscript{2},
        \textbf{Johann Heuser}\textsuperscript{2},
        \textbf{Joerg Lehnert}\textsuperscript{2},
        \textbf{Christian Joachim Schmidt}\textsuperscript{2},
        \textbf{and}
        \textbf{Dennis M\"ucher}\textsuperscript{1}
    \end{center}
    
    \textsuperscript{1}Institute for Nuclear Physics, University of Cologne, 50937 Cologne, Germany
    
    \textsuperscript{2}GSI GmbH, 64291 Darmstadt, Germany
    
    \section{Abstract}
    \label{sec:rate_abstract}
    
    Designed to provide quality assurance for ion-beam radiotherapy, the prototype fIVI (filtered Interaction Vertex Imaging) Range Monitoring System is a two-layer tracker which employs double-sided strip-segmented silicon detectors. To meet the high demands of a clinical environment, a large sensitive area is required, along with a fast and compact readout. As this device utilizes sensors and readout electronics adapted from particle physics, where the expected energy and count rate differ significantly from radiotherapy, validation was necessary to ensure that these sensors would function effectively at the order \qty{100}{\MeV\per\nucleon} energies and order \unit{\mega\hertz} count rates expected during clinical irradiation. Tests were conducted using scattered subclinical \qty{19}{\MeV} protons at high intensity, and clinical \qty{207}{\MeV\per\nucleon} carbon ions at low intensity to independently validate these variables. The detection system is found to operate at rates up to \qty{1.3}{\mega\hertz}, with a negligible fraction of events being affected by pileup. The efficiency of hit reconstruction is high, with a timestamp resolution of \qty{6.25}{\nano\s}, and a coincidence window of \qty{31.25}{\nano\s}, as is required for clinical event rates. With these settings, over \qty{90}{\percent} of particle interactions are able to reconstruct unique hit positions and contribute to track formation. This device is the first system using large-area, high-resolution detectors which meets the demanding count rate requirements associated with clinical radiotherapy.
    
    \section{Introduction}
    \label{sec:rate_intro}
    
    A growing modality for external-beam radiation therapy is Carbon Ion Radiotherapy (CIRT), which uses a \cion{} beam to deliver dose to the tumour \autocite{malouff_carbon_2020}. Unlike the more common photon-based radiotherapy, the charged beam loses energy quasi-continuously in matter, resulting in a beam which slows and stops within the patient. The dose distribution resulting from this behaviour is inverted relative to that of a photon beam, resulting in the dose maximum, or Bragg peak (BP), being located where the beam stops \autocite{amaldi_radiotherapy_2005}. Therefore, CIRT is capable of delivering highly precise conformal dose distributions, maintaining a high dose to the tumour while sparing the surrounding healthy tissue \autocite{amaldi_radiotherapy_2005}. As most tumours are significantly larger than can be uniformly irradiated by a single BP, CIRT treatments commonly involve a constellation of BP positions and intensities \autocite{kramer_treatment_2000}. Full tumour coverage is achieved by varying the lateral position through magnetic steering of the beam, and varying the BP depth by changing beam energy \autocite{haberer_magnetic_1993}. As with most forms of external beam radiation therapy, treatments are also commonly fractionated. In a fractionated treatment, the total prescribed dose is delivered over several irradiations, usually no more than once per day over a period of weeks. The separation between fractions allows cellular repair mechanisms in healthy tissue to moderate the damage to non-cancerous cells \autocite{amaldi_radiotherapy_2005}.
    
    However, the dosimetric advantage of CIRT also introduces the additional challenge of monitoring the Bragg peak position, to ensure it is located directly on the tumour. Currently, safety margins on the range of millimeters are required around a tumour, to account for uncertainties in beam delivery and stopping range \autocite{kelleter_-vivo_2024, andreo_clinical_2009}. Range monitoring (RM) methods are used to measure the BP depth, with the goal of improving the consistency of dose delivery, and reducing the required safety margins \autocite{finck_study_2017, hymers_intra-_2021}. Achieving these goals would allow a reduction in the dose delivered to healthy tissue surrounding the tumour, while maintaining the currently achievable level of tumour control.
    
    One RM method which is currently undergoing clinical trials for use in CIRT is Interaction Vertex Imaging (IVI) \autocite{fischetti_inter-fractional_2020, kelleter_-vivo_2024}, which uses external detectors to monitor prompt secondary charged particles (primarily protons) created by beam-patient reactions \autocite{amaldi_advanced_2010}. These external detectors track the particle, and backproject its trajectory into the patient to approximate the location of the reaction (or interaction vertex) which produced it \autocite{finck_study_2017, henriquet_interaction_2012, hymers_intra-_2021}. However, the number of particles which can be reconstructed is limited, with this limit directly proportional to the number of primary \cion{} ions delivered to each BP \autocite{finck_study_2017, henriquet_interaction_2012}. The number of primary ions delivered is in turn directly determined by the prescribed dose and fractionation scheme \autocite{kramer_treatment_2000, amaldi_radiotherapy_2005}. Therefore, to facilitate IVI RM, it is necessary to collect as many of these secondary particles as possible. Unlike in pure nuclear or particle physics applications, it is not possible to extend the measurement time in order to collect additional statistics, as this extension would result in a departure from the prescribed dose.
    
    In addition to the requirement for high collection efficiency, it is also necessary that this efficiency be achieved at high count rates. Typical CIRT beam intensities range from \qtyrange{e6}{e8}{\ion\per\s} to achieve target dose rates on the order of \qty{1}{\gray\liter\per\minute} \autocite{kramer_treatment_2000}, with significantly higher beam intensities of \qtyrange{e9}{e12}{\ion\per\s} required for ultra high dose-rate FLASH radiotherapy \autocite{schoemers_christian_beam_2023}. In the conventional radiotherapy case, higher dose rates facilitate treatment of larger tumours, while in the case of FLASH radiotherapy, the higher dose rates result in further sparing of healthy tissue; the exact mechanism of this protective effect is still debated \autocite{hughes_flash_2020}. Regardless of the irradiation modality, it is necessary that any IVI RM hardware be capable of handling at least the maximum conventional dose rate, to not negatively affect the existing patient experience \autocite{hymers_intra-_2021}. The further ability of the RM hardware to handle FLASH dose rates would provide the additional benefit of monitoring these novel and demanding treatment plans.
    
    Therefore, the ideal system for IVI RM requires a large sensitive area, to maximize the geometric efficiency of data collection, combined with a data acquisition system capable of handling high input rates without significant losses due to pileup \autocite{finck_study_2017, hymers_intra-_2021}. The application of state-of-the-art detector technology from particle physics facilitates high-rate readout. One approach, currently being studied by a group at the Heidelberg Ion-Beam Therapy Centre (HIT), is to combine many smaller devices into a tracker array, resulting in a larger aggregate sensitive area \autocite{kelleter_-vivo_2024}. The resulting array is highly capable, but is limited by the larger structure to certain patient and tumour configurations, and requires a large number of sensors to be synchronized and read out, increasing the complexity of the system. Another approach is to use larger sensors which, although potentially limiting in overall count rate, are believed to provide sufficient performance for at least conventional irradiation \autocite{fischetti_inter-fractional_2020, hymers_intra-_2021}.
    
    This work describes a single tracker, comprised of fast, large-area silicon sensors, and investigates its suitability for IVI RM. The acquisition and tracking performance of this device, demonstrated at highly nonuniform event rates up to \qty{1.3}{\mega\hertz}, compares very favourably to the requirements of conventional and FLASH radiotherapy.
    
    \section{Materials and Methods}
    \label{sec:rate_materials}
    
    \subsection{CBM STS Sensor}
    \label{sec:rate_sensor}
    
    The large-area sensors used in the prototype fIVI Range Monitoring System were originally developed for the Silicon Tracking System of the Compressed Baryonic Matter experiment at GSI \autocite{heuser_technical_2013}. For range monitoring, the largest sensors, with sensitive areas of \qtyproduct[product-units=power]{60x60}{\mm} and \qtyproduct[product-units=power]{120x60}{\mm} and a \qty{300}{\micro\m} thickness were selected, to maximize solid angle coverage with only a single sensor per layer. Position sensitivity was achieved through segmentation of both sides of the planar sensor into 1024 segments per side, at a \qty{58.6}{\micro\m} pitch. On the n side of the sensor, these segments were oriented axially, while on the p side of the sensor, these segments were angled at \ang{7.5} from the axis. To achieve full coverage on the p side without expanding the number of segments beyond 1024, segments which reach the lateral edge of the sensor were connected to a corresponding segment on the opposite edge using a second metallization layer. The result of this connection was a Z-shape, electrically connecting two parallel sensitive regions as shown in \figref{segmentation}, and the affected segments were termed `Z-strips'. 
    
    \doubleFig[!t]{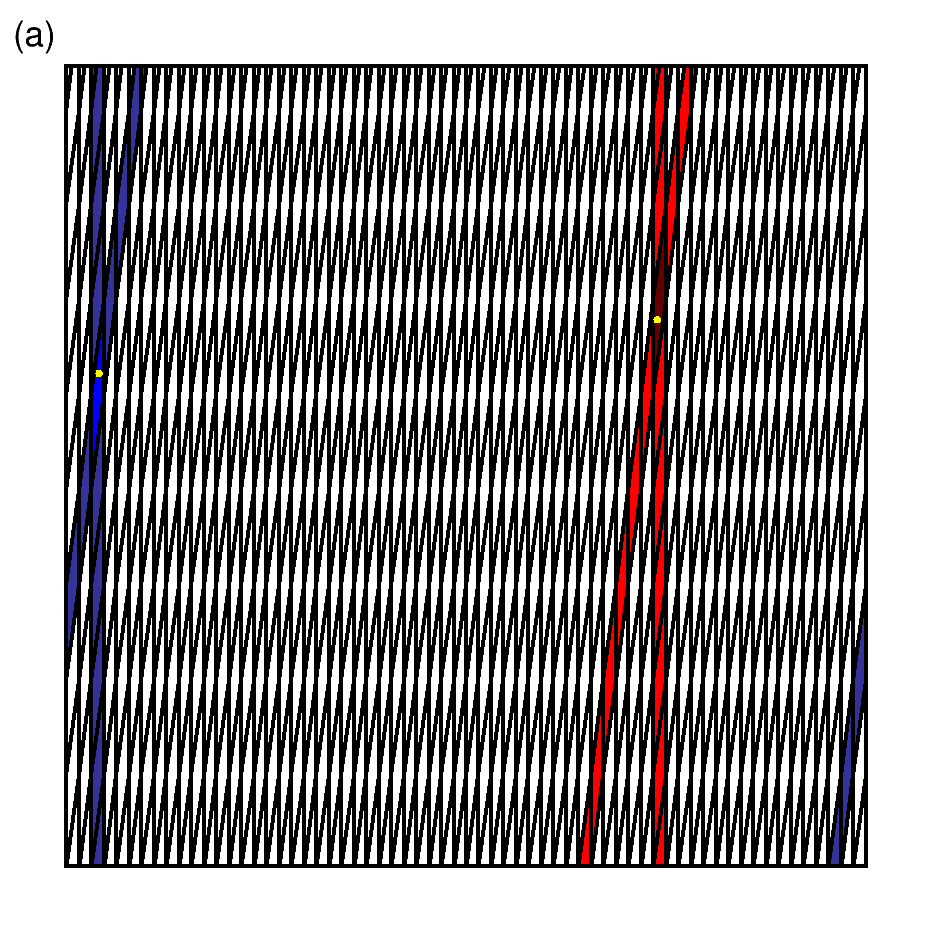}{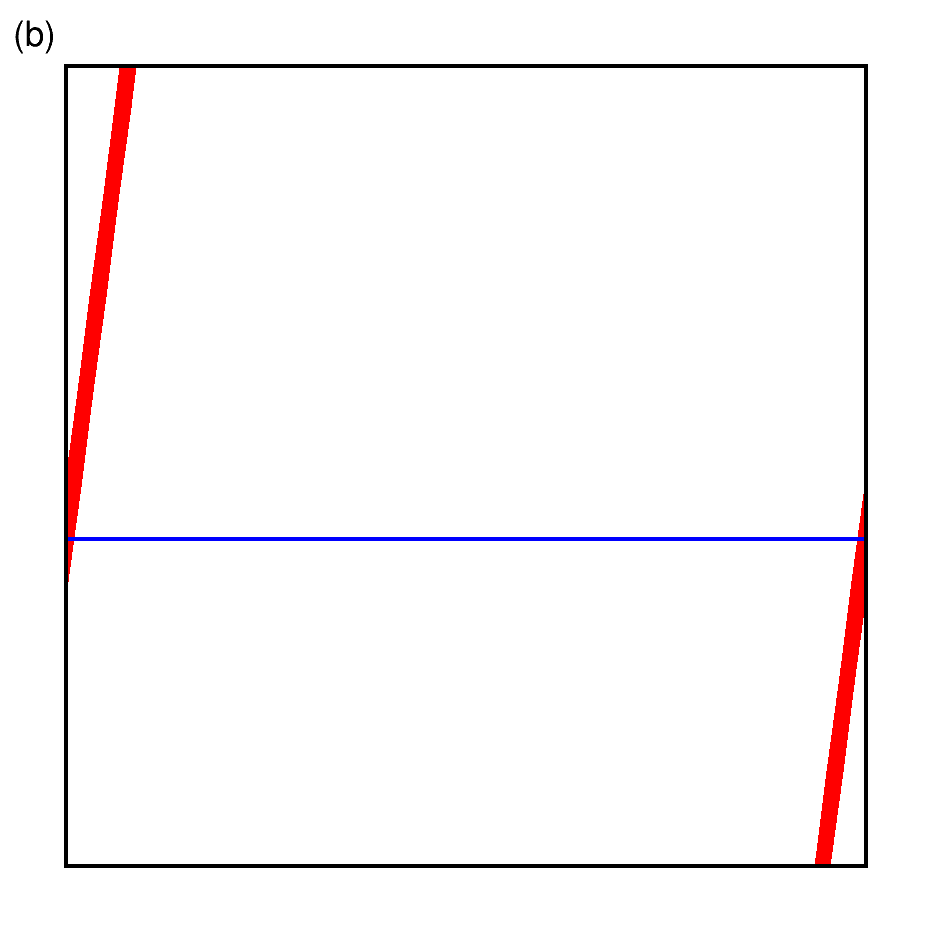}{segmentation}{(a) Representative schematic of a sensor, showing the kite-shaped elements formed by the intersection of axial and angled segments. Two interaction positions are shown (yellow), along with the active segments (blue, red). Note the Z-strip angled segment (blue) which wraps around the lateral edge of the sensor. Due to the use of angled segments, there is no confusion between the two hits. In the case of orthogonal segments, these two interaction positions would instead produce four candidate positions, as each segment on one side intersects with every segment on the opposite side. (b) Detail view of a Z-strip segment, with the two disjoint sensitive regions (red) joined by an electrical interconnect (blue). This connection allows the entire sensor to be read out along the bottom edge, while maintaining position sensitivity across the entire sensor area.}{Representative schematic of a sensor.}
    
    The combination of angular and axial strips allowed all electronic connections to be made along a single \qty{60}{\mm} edge of the sensor, facilitating the design of larger detector arrays consisting of adjacent sensors. However, this design decision reduced the position resolution as compared to the more typical design of orthogonal strips on opposite sensor sides. In particular, resolution along the length of the sensor was significantly reduced: the angled strips elongated each pixel (as defined by a unique intersection of a single p-side channel with a single n-side channel) to a center-to-center distance of \qty{445}{\micro\m} in length, while retaining a width defined by the segment pitch of \qty{58.6}{\micro\m}. Consequently, the spatial resolution of the sensors was highly orientation-dependent. Another advantage of these angled strips was the reduction in ghost hits formed by the coincident interaction of multiple particles with the sensor, also seen in \figref{segmentation}.
    
    \subsection{Tracker Design}
    \label{sec:rate_tracker}
    
    As range monitoring using IVI relies on particle tracking, a minimum of two sensor layers were required. While additional sensor layers could provide redundancy, and an ability to reject particles which scattered within an early sensor layer, the probability of significantly scattering protons with energies of \num{100}s of \unit{\mega\eV} in the \qty{300}{\micro\m} silicon was found to be sufficiently low as to not warrant the expense and additional complexity of the extra readout channels. Therefore, the final tracker consisted of two sensors, a front layer with a \qtyproduct[product-units=power]{60x60}{\mm} sensitive area, and a rear layer with a \qtyproduct[product-units=power]{60x120}{\mm} sensitive area. Using a longer rear sensor allowed for a larger effective field of view, and correspondingly greater detector acceptance, which is of critical importance when the total number of primary particles is fixed by a radiotherapy treatment plan.
    
    Each sensor was mounted to a thin PEEK (polyethyl ether ketone) plate, with a cutout for the sensor’s sensitive area. The plate and sensor assembly were then mounted in a custom-designed aluminum carrier frame, which housed the sensor and all readout ASICs for both sides. This entire assembly of sensor, supports, and frame was termed a `module'; a complete module can be seen in \figref{module12}. The frame allowed the sensor to be mounted in various configurations, while maintaining the appropriate immobilization of shielding layers around the microcables connecting the sensor to the readout electronics. Provision was also made for attaching a cold plate heat exchanger directly to the readout electronics mount, in support of maintaining proper operating temperature.
    
    \singleFig[t]{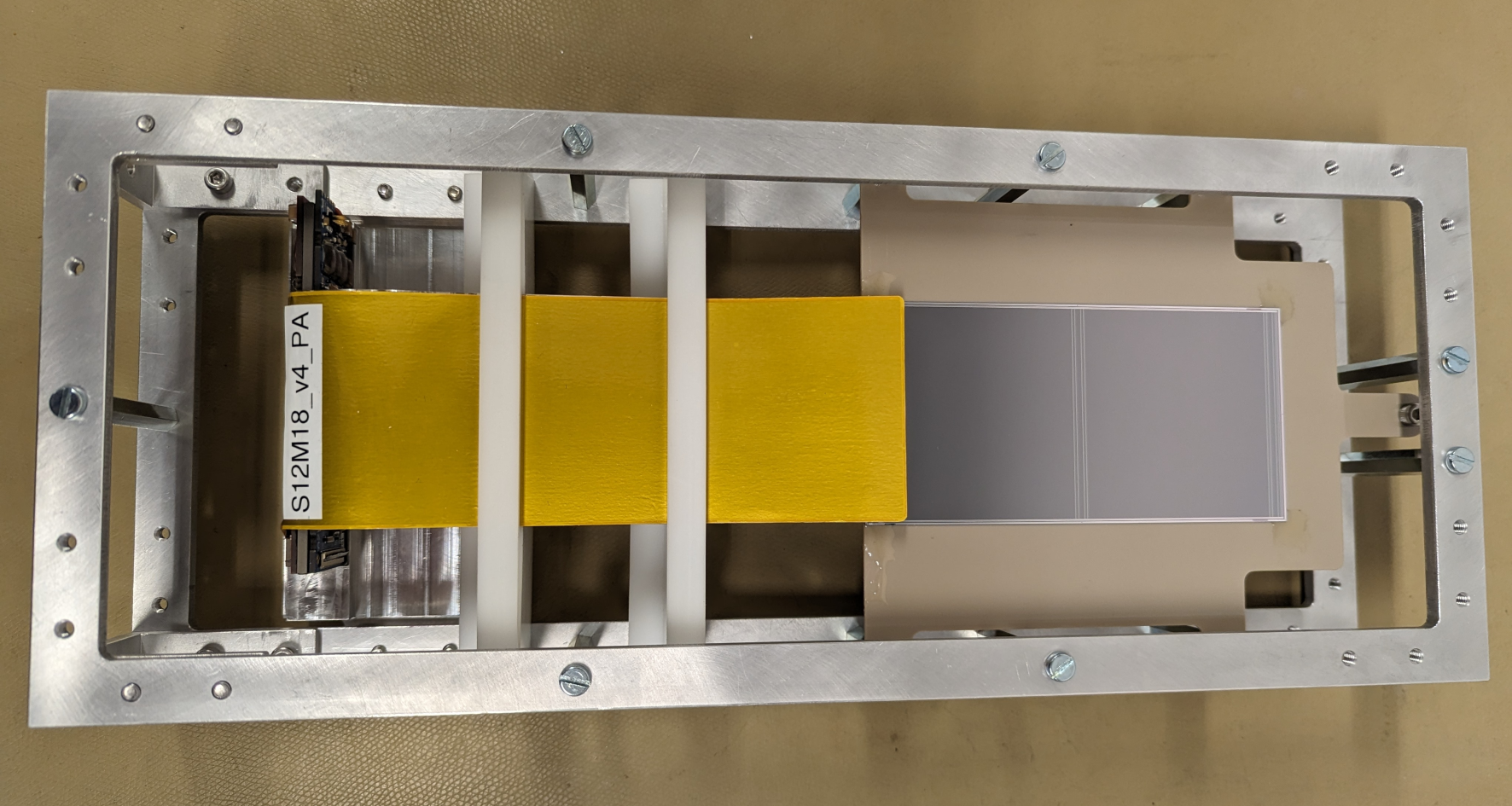}{module12}{Complete \qtyproduct[product-units=power]{60x120}{\mm} module. The sensor (right) was affixed to the backing at its four corners. Only the \qty{1}{\mm} margin of the sensor guard ring was directly supported by the backing; the portion corresponding to the sensitive area was removed. The sensor and backing were mounted \qty{35}{\mm} in front of the rear frame. A shielded bundle of analog microcables (center, yellow) connected the sensor and the front end electronics (left). These cables were held in place by two clamps (white). The front frame was separated from the rear frame by \qty{50}{\mm}.}{Complete \qty{60}{} $\times$ \qty{120}{\square\mm} module.}
    
    To form a tracker setup, the two modules were fixed to an aluminum baseplate such that their sensitive surfaces were parallel, separated by \qty{12}{\cm}. A protective aluminum enclosure was built atop this baseplate, as shown in \figref{tracker}, to isolate the sensors from visible light. The walls of this enclosure were \qty{1.0}{\mm} thick, with a thinner \qty{16}{\micro\m} aluminum entrance window placed parallel to the front sensor. The distance between this window and the front sensor was \qty{3.35}{\cm}. This entrance window was sufficiently large as to not limit the effective field of view of the tracker. Electrical connections for power and data passed through the enclosure using panel-mount connectors, while bulkhead feedthroughs allowed liquid cooling.
    
    \doubleFig[!t]{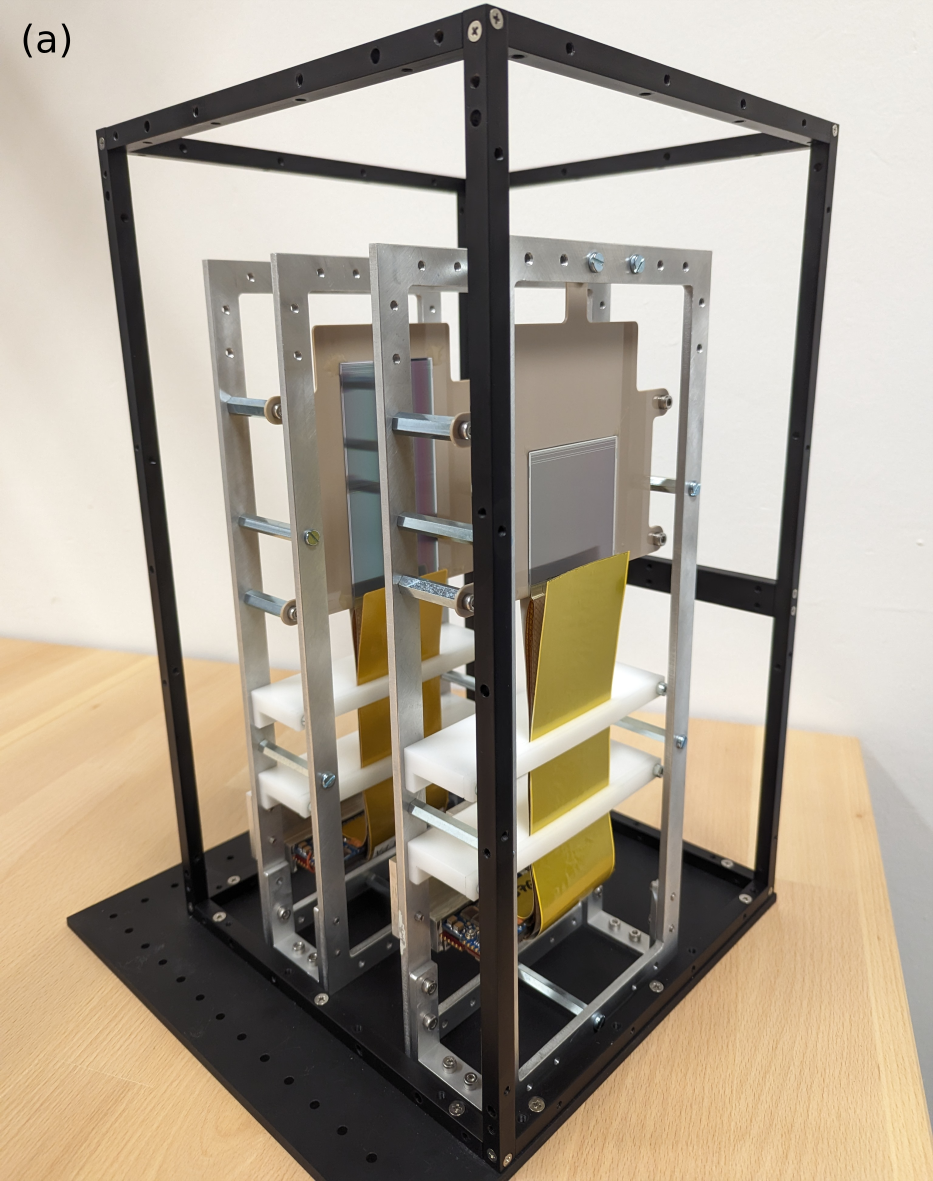}{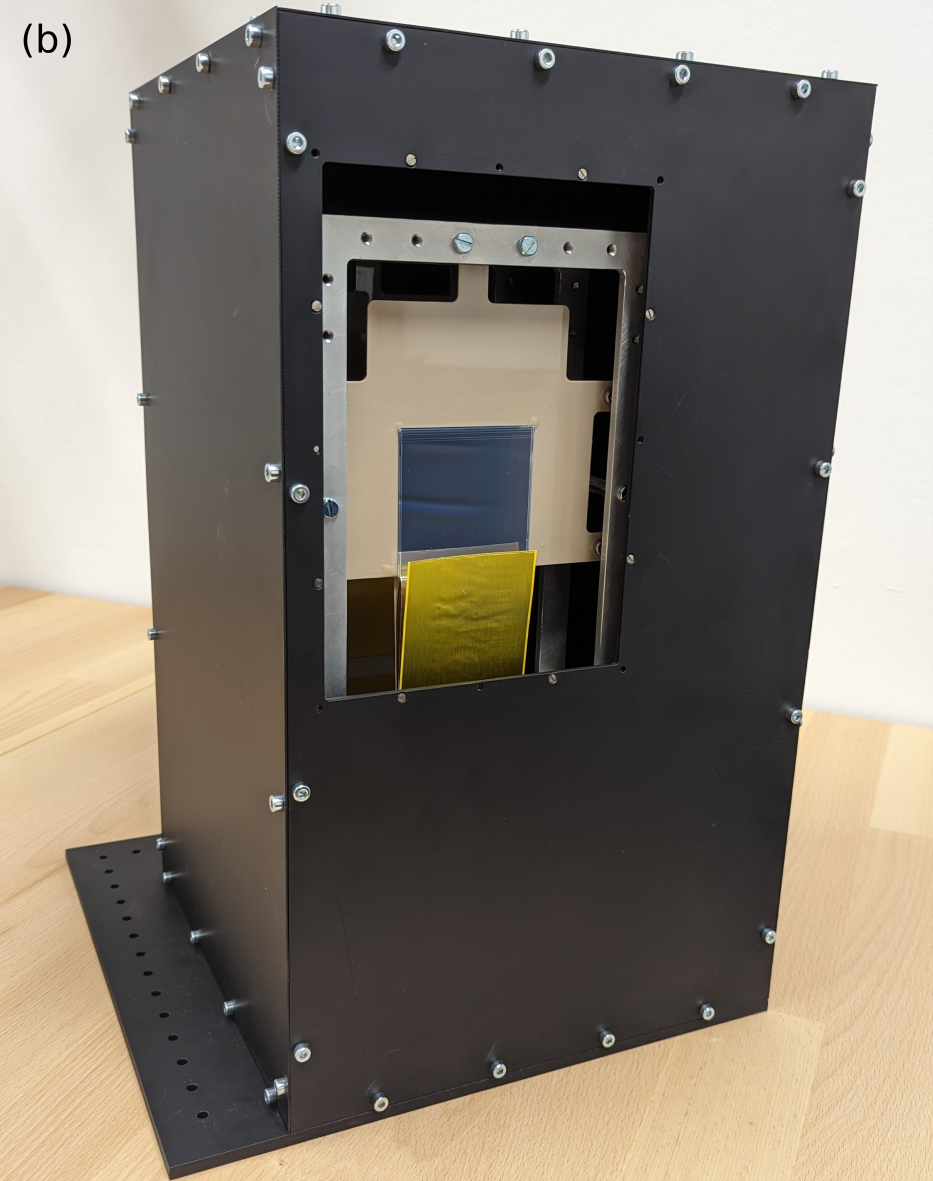}{tracker}{Both modules installed in the tracker. (a) Covering panels removed, showing the modules mounted to the base plate, with the sensors separated by \qty{12}{\centi\m}. (b) Covering panels installed, with the exception of the entrance window. For data collection, this window region was covered with a \qty{16}{\micro\m} thick aluminum foil. The region of the base plate exterior to the enclosure was used to affix the tracker to the experimental setup.}{Both modules installed in the tracker.}
    
    \subsection{Front End Electronics}
    \label{sec:online_front_end}
    
    The core electronic component of the fIVI Range Monitoring System is the SMX 2.2 application-specific integrated circuit (ASIC) \autocite{kasinski_sts-xyter_2014, kasinski_back-end_2016}. Each chip provides fast and compact digital readout of up to \num{128} analog channels, with a 14-bit, \qty{6.25}{\nano\s} precision timestamp, dynamic range up to \qty{100}{\femto\coulomb} in its low-gain operation mode, and a 5-bit analog-to-digital converter (ADC) for energy measurement using pulse height analysis. Fast timing is provided by splitting the input signal to two separate shaping circuits: a `fast’ shaper with a peaking time on the order of nanoseconds to provide timestamp information; and a `slow’ shaper with a peaking time of \qty{90}{\nano\s} which provides higher precision energy information \autocite{kasinski_front-end_2016}. Although the energy resolution performance is reduced when compared to a longer shaping time, this tradeoff is made to allow faster reset of the readout electronics, which reduces dead time. Each of the \num{128} channels on the ASIC can have individual thresholds set, or even be entirely disabled, to achieve uniform sensitivity between channels. A one-time calibration process is required after the ASIC is configured, which can then be saved and reapplied by the control PC when the data acquisition system is powered on \autocite{saini_test_2023}.
    
    \singleFig[!t]{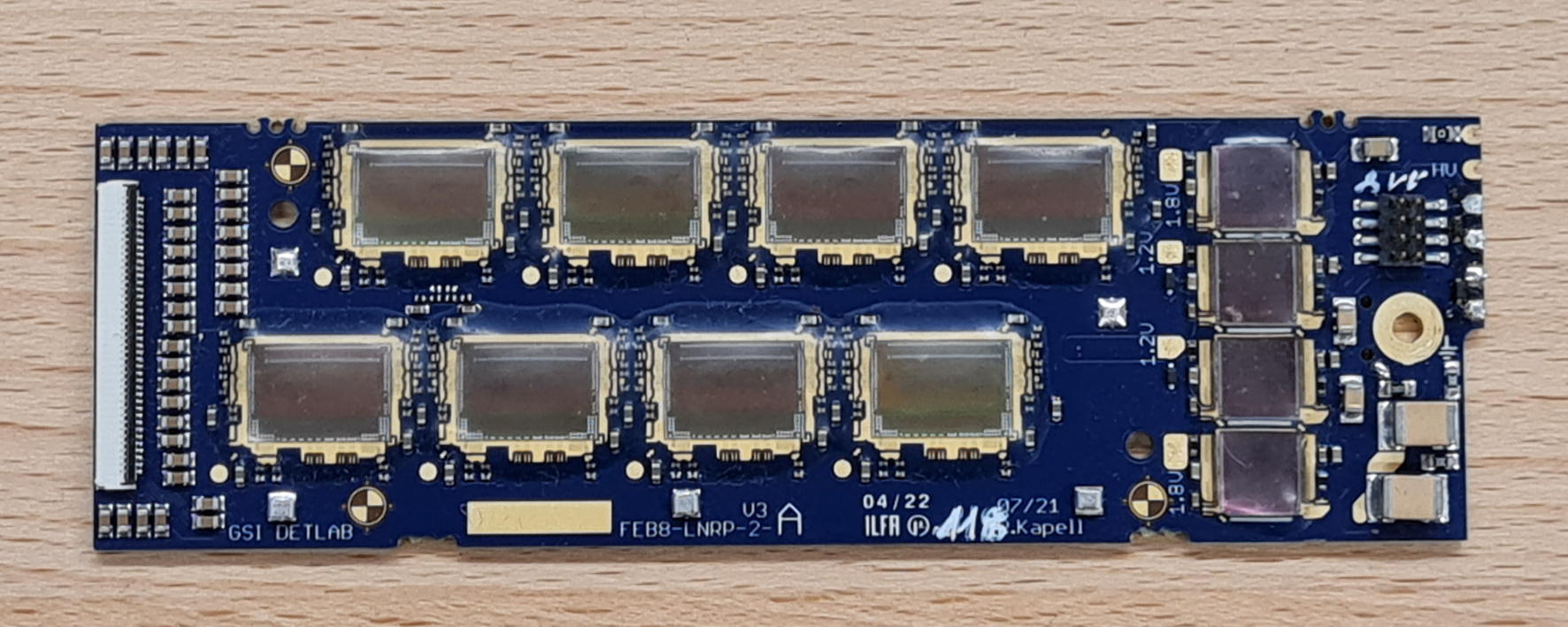}{feb}{Front-end board used for readout of one side of a silicon sensor. Eight SMX ASICs are mounted in two staggered rows on the center of the board. Communication occurs through a ribbon cable inserted into the socket on the left, which provides a common clock distributed to all ASICs, a common control interface which can address commands to an individual ASIC or to all ASICs simultaneously, and 16 independent data uplinks (two per ASIC). The eight-pin header on the upper right provides low-voltage power for the analog and digital electronics on each ASIC, as well as a bias voltage for the connected sensor.}{Front-end board used for readout of one side of a silicon sensor.}
    
    As the silicon sensors used in the Range Monitoring System are double-sided strip-segmented detectors, with \num{1024} segments per side, eight SMX ASICs are required to read out each sensor side. These eight ASICs are mounted in a shared printed circuit board called a `Front-End Board' (FEB), pictured in \figref{feb}. The FEB also contains voltage regulators for low-voltage ASIC power, operating with the negative terminal floating on the bias voltage applied to the given sensor side. Sensor segments are directly connected to the SMX ASICs using analog microcables as shown in \figref{module12}, with lengths between \qty{149}{\mm} and \qty{180}{\mm}. Different cable lengths are used to facilitate the different sensor lengths, with shorter cables used on the larger sensor to allow similar positioning of the readout electronics relative to the sensor. Slight differences also exist between the cables for the p and n sides of the sensor, to account for an \qty{11}{\mm} difference in path length from the bottom edge of the sensor to each set of SMX ASICs. Each cable, as well as the overall cable bundle, is shielded to reduce the impact of electromagnetic noise on sensor performance. Although the SMX ASIC is radiation-resistant \autocite{kasinski_characterization_2018}, using long cables allowed the readout electronics to be placed out of the direct radiation field, and facilitated the use of liquid cooling to dissipate the up to \qty{50}{\watt} produced by the low-voltage power and readout electronics of each sensor.
    
    Communication with these ASICs occurs over the GBTx-compliant elink physical protocol, using low-voltage differential signalling \autocite{moreira_gbtx_2021}. Clock distribution and downlink for control signals are implemented using a multi-drop link, with a single driver and eight receivers, one per ASIC. This topology allows the addressing of commands to individual ASICs using their unique addresses, or to a broadcast address which issues commands to all ASICs on the same FEB. Each ASIC also supports two independent uplinks, for sending data concerning digitized sensor interactions to be passed to the data acquisition system. All data links operate with AC-coupled 8b/10b-encoded signalling on the shared \qty{80}{\mega\Hz} clock; the uplinks operate at dual data rate to transmit at an elevated \qty{160}{\mega\bit\per\second} \autocite{zabolotny_gbtx_2021-1}.
    
    Communication over these digital data links uses a custom protocol with a constant 24-bit frame size. This protocol enables partial compression of data by transmitting the six most significant bits of the 14-bit timestamp separately, in a special `TS\_MSB' frame. These six bits are then applied to all trigger events, represented by `Hit' frames, until the next TS\_MSB frame is sent. This optimization allows lossless compression, and results in lower total data transmission whenever four or more trigger events transmitted over the same uplink share the six most significant timestamp bits, a scenario expected to occur at high data rates \autocite{kasinski_protocol_2016}. Each uplink separately tracks the most recent TS\_MSB bits, and generates a new frame whenever the next Hit frame to be sent has a different set of most significant bits. To account for periods of time with low event rates, special TS\_MSB frames are also sent whenever the five most significant timestamp bits change, so long as no Hit frames are ready to be sent. These frames provide later parts of the readout and data processing system with information about the passage of time, avoiding the possibility of an error in global timestamp assignment for two consecutive events on the same ASIC separated by more than \qty{102.4}{\micro\s}, the period of the 14-bit timer \autocite{kasinski_protocol_2016}.
    
    \subsection{Back End Electronics}
    \label{sec:online_back_end}
    
    \singleFig[!b]{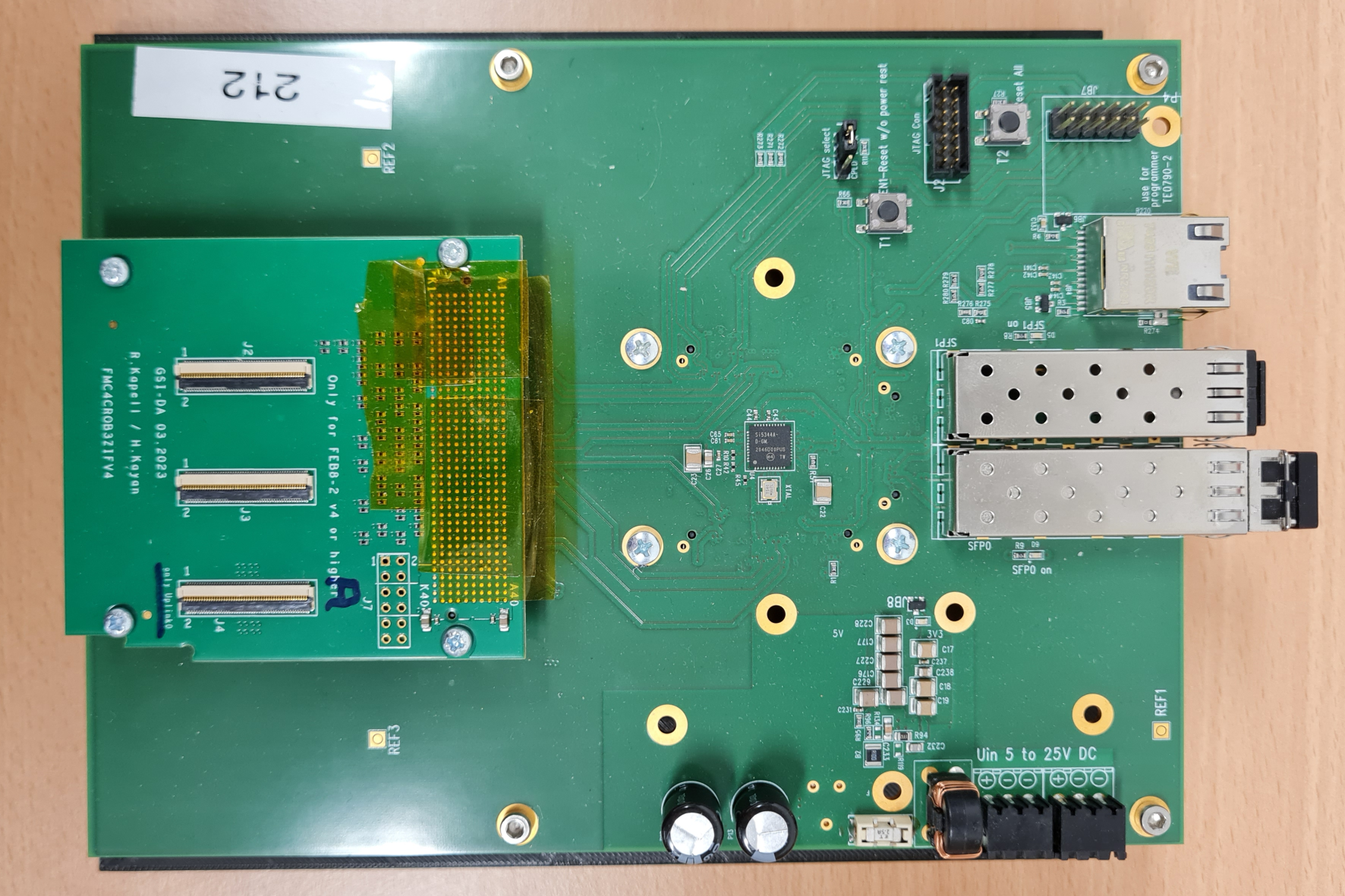}{emu}{GBTX Emulator board. Connections to front-end electronics are made via ribbon cables connected to the three sockets on the left. Only the upper two sockets are active during data collection, with the uppermost socket supporting 16 simultaneous uplinks, and the middle socket being restricted to 12. The Artix 7 FPGA is mounted to the underside of the board, in the center. Power is provided through terminals on the lower right edge of the board. Upstream interfaces are visible on the right, with the active optical link in the lower cage. The second optical link and the ethernet interface for slow control are not used in this setup.}{GBTX Emulator board.}
    
    Communication with one or more FEBs is handled by the AMD/Xilinx Artix 7-based field-programmable gate array (FPGA) GBTX Emulator board (GBTxEMU), pictured in \figref{emu} \autocite{zabolotny_gbtx_2021-1}. This board is an emulator for the core functionality of the GBTX ASIC developed at CERN for standardized data control and readout, and is suitable for low-cost applications where the full radiation hardness of the GBTX is not required. For the fIVI Range Monitoring System, the lower cost of the GBTxEMU was attractive, and the ability to place these readout boards outside of the primary radiation field, and away from the forward-focused secondary fragments, made radiation hardness a secondary consideration. Each readout board aggregates data from up to \num{28} uplinks operating at \qty{80}{\mega\Hz}, and directly retransmits this data along an optical GBT uplink to the data processing board. The optical uplink operates at a frame frequency of \qty{40}{\mega\Hz} in GBTX wide frame mode, with a total available bandwidth of \qty{4.48}{\giga\bit\per\second}; this bandwidth is what limits the number of uplinks per readout board to \num{28} \autocite{zabolotny_gbtx_2021}. Control commands for individual SMX ASICs are also delivered to the GBTxEMU via a corresponding optical downlink; for communication integrity, this link operates in the standard GBT frame format, with a reduced bandwidth of \qty{3.36}{\giga\bit\per\second}. The remaining raw link bandwidth is used for error correction.
    
    Data from all readout boards is aggregated in the data processing board, termed the GBTxEMU Readout Interface (GERI) \autocite{zabolotny_versatile_2023}. This board is a PCI Express 3.0 8-lane add-in card for a standard workstation PC, with an AMD/Xilinx Virtix-7 FPGA providing onboard processing. The role of the GERI is to manage the optical links for up to seven GBTxEMU boards, handling broadcast and individual board commands, as well as receiving the wide frame uplink data transmitted from each SMX elink. The GERI reverses the 8b/10b encoding of the SMX data frames, eliminates synchronization and empty frames, tags each frame with the appropriate topology (using a 3-bit GBTxEMU ID and 5-bit elink ID), and delivers the ordered, tagged data to the main system memory over the PCI Express bus, through direct memory access (DMA). This board also provides the reference clock used to drive the optical interfaces and elinks. The GERI driver software provides two separate interfaces: a control interface, through which commands can be issued to one or more SMX ASICs; and the data acquisition interface, which handles the delivery of DMA data and communication with the data processing code running on the workstation PC.
    
    In the current setup, each sensor in the fIVI Range Monitoring System is read out by two FEBs, one per sensor side, for a total of four boards and \num{32} SMX ASICs. At full capacity, these sensors are capable of utilizing \num{64} elink uplinks, requiring bandwidth corresponding to at least three GBTxEMU boards. Due to hardware limitations of the current setup which require each FEB be connected to only one GBTxEMU board, four readout boards would be required for full link utilization. However, it is possible to divide the \num{28} elinks of a single GBTxEMU between two FEBs, such that all ASICs on one FEB are allowed two uplinks, and all ASICs on the second FEB are allowed at least one uplink, with four of the eight having both uplinks active. Both of these GBTxEMU boards may then be controlled and read out by a single GERI.
    
    \subsection{Simulation Detail}
    \label{sec:rate_sim}
    
    The tracker setup was implemented in the Geant4 Monte Carlo simulation toolkit \autocite{agostinelli_geant4simulation_2003}, version 11.1. This simulation was used to evaluate the required performance of the tracker for clinical applications, as well as to provide a reference to compare to experimental results.
    
    To model the segmentation of the sensors, each sensor was split into pixels, with each pixel representing the intersection of a p side and an n side segment. Interactions in each pixel contributed to both corresponding segments, and only the segment information was made available for analysis. In addition to the segment emulation, the exact interaction position was also recorded in a separate data structure. This division allowed two forms of analysis: one which made use of only the data available by the segmented system, and one which allowed consideration of metadata which was only available in the simulation.
    
    \doubleFig[!b]{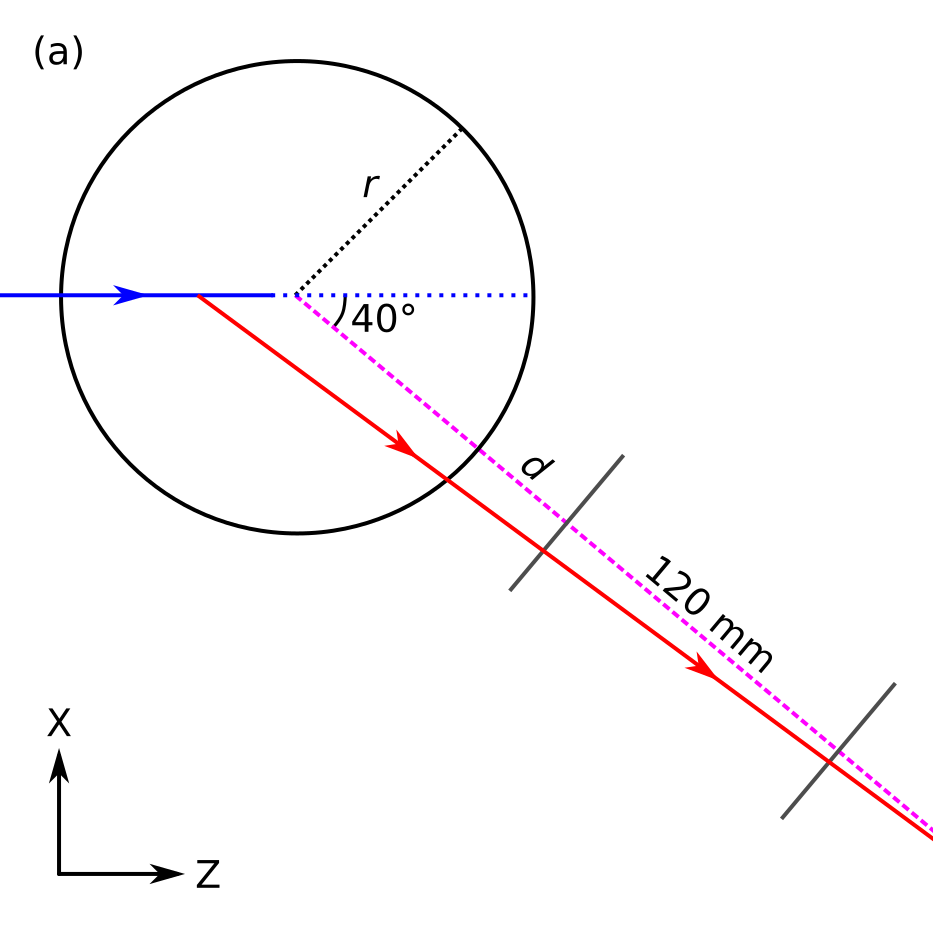}{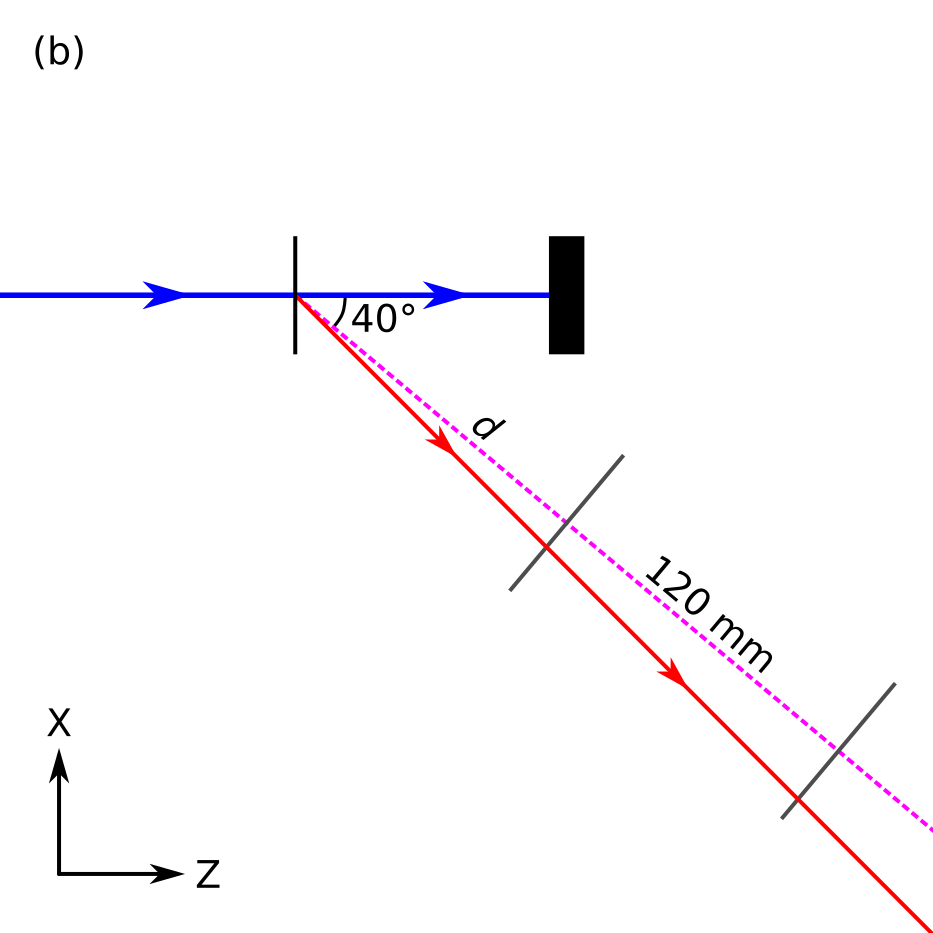}{setup_sim}{Schematics of simulation geometries. (a) Model of clinical irradiation of a cylindrical phantom, with radius $r = \qty{10}{\cm}$. The treatment beam (blue) interacts with the phantom, producing secondary particles (red), which may be recorded by silicon sensors (grey). (b) Model of scattering from target foil. The primary beam (blue) passes through the thin target foil (black), before stopping in the beam dump (black, thick). Scattered particles (red) may interact with silicon sensors (grey). In both geometries, the central axis of the fIVI Range Monitoring System (magenta, dotted) is placed at a \ang{40} angle from the primary beam axis (blue). The separation $d$ between the target and the first sensor may be configured within each setup.}{Schematics of simulation geometries.}
    
    As the distribution of secondary particles used in IVI is strongly dependent on angle \autocite{finck_study_2017, ghesquiere-dierickx_investigation_2021}, a reaction producing a similar angular distribution was sought, to allow investigation of the performance of the CBM sensors under nonuniform irradiation. While Rutherford scattering produces a distribution which is more strongly dependent on angle than the IVI secondary particles, it was deemed an acceptable surrogate. The simulated tracker was placed at a \ang{40} angle from the primary beam axis. Two configurations, both shown in \figref{setup_sim}, were simulated. The first configuration modelled a clinical \cion{} beam in a \qty{20}{\cm} diameter plastic phantom, while the second modelled a low-energy proton beam scattering on a \qty{100}{\micro\m} thick target. All simulations used the QGSP\_BIC physics list, which has been previously demonstrated to suitably reproduce the angular distribution of secondary particles for beam energies below \qty{200}{\mega\eV\per\nucleon} \autocite{finck_study_2017}.
    
    Simulations completed with the plastic phantom suggested that, at primary beam intensities up to \qty{1e9}{\ion\per\s}, the front sensor of the tracker would be expected to experience an event rate of approximately \qty{1.2}{\mega\hertz}. As this primary beam intensity is a factor of 10 higher than the typical beam current used in CIRT, and the QGSP\_BIC physics list is known to overestimate secondary particle yields by a factor of two \autocite{gwosch_non-invasive_2013}, this count rate was adopted as an appropriate benchmark that the detection system should be able to handle without significant losses due to pileup or dead time. Although this estimate was significantly higher than the interaction rate which would actually be expected in a clinical configuration, this was a purposeful overestimate. Overestimating the hit rate at low beam energy was chosen to allow for larger cluster sizes at high beam energy, which might result from multiple sensor segments being triggered by the same particle. This overestimate also accounted for a possible longer segment reset time due to the higher energy deposit in the sensor from a clinical beam, leading to longer dead times.
    
    \subsection{Low-Energy Test}
    \label{sec:rate_le}
    
    To test the performance of the system at clinically-relevant data rates, and with both sensors active, a beam test was performed at the Cologne \qty{10}{\mega\volt} FN Tandem accelerator. Primary protons were accelerated to an energy of \qty{19}{\mega\eV}, and scattered on a \qty{100}{\micro\m} aluminum foil. The fIVI Range Monitoring System was placed at a \ang{40} angle, with this foil centered in its field of view, and with a \qty{186.4}{\mm} distance between the center of the foil and the front sensor. The installed setup is shown in \figref{setup_le}. Both sensors were symmetrically biased at \qty{\pm 70}{\volt}, for a total bias voltage of \qty{140}{\volt}.
    
    \singleFig[!t]{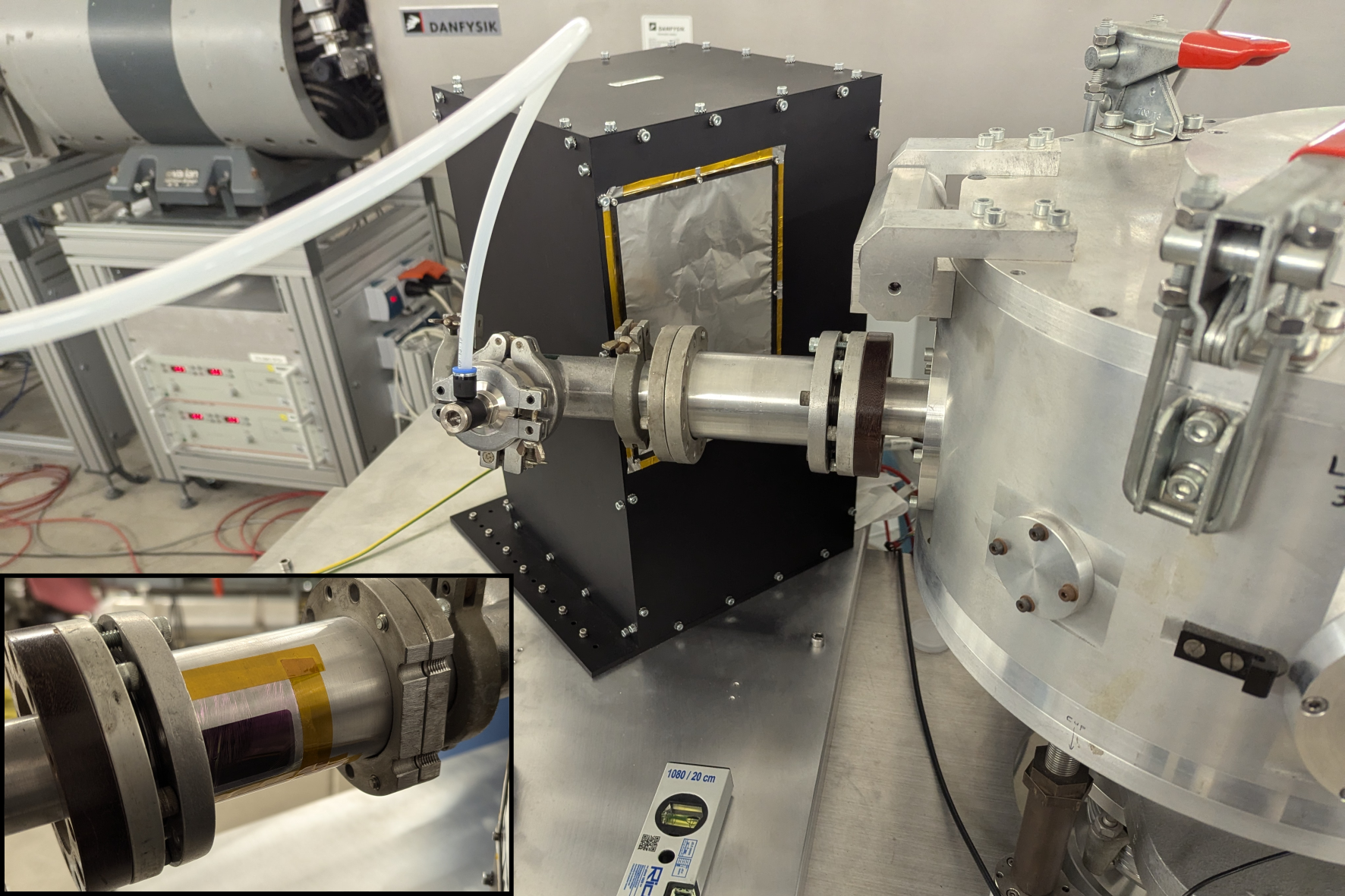}{setup_le}{Experimental setup for low-energy rate test. The \qty{19}{\mega\eV} proton beam entered the setup from the right, and was incident on a \qty{100}{\micro\m} aluminum foil, which acted as both the target and the vacuum window. Scattered protons from this foil passed through the \qty{0.5}{\micro\m} mylar foil covering a cutout of the beam pipe (inset), and passed through air before reaching the tracker. The tracker was placed at a \ang{40} angle from the beam axis, as smaller angles would interfere with the pipe containing the beam dump. The first sensor was placed \qty{186.4}{\mm} from the center of the target foil, with the second sensor a further \qty{120}{\mm} away. The exterior beam pipe was continually flushed with helium to limit interactions of the primary beam beyond the window, before stopping in the aluminum beam dump.}{Experimental setup for low-energy rate test.}
    
    The kinematic calculator of LISEcute++ \autocite{tarasov_lise_2016} was used to estimate the proton beam intensity which would correspond to the target event rate of \qty{1}{\mega\hertz}, yielding an intensity of \qty{50}{\nano\ampere}. However, the target event rate was reached at lower beam intensities, due to multiple scattering on other setup elements, and additional contributions from non-Rutherford processes. A variety of primary beam intensities up to \qty{20}{\nano\ampere} were delivered for \qty{60}{\s} irradiations, along with beam-off background measurements. Due to beam instability, the exact intensity could only be estimated from Faraday cup measurements performed immediately before and after beam on target.
    
    Two configurations were tested, the first with a single uplink active for each SMX ASIC, and the second with all possible uplinks active. Due to previously-discussed bandwidth limitations, this all-uplink configuration resulted in two uplinks per SMX being active for all eight SMXs on the n side of each sensor and half of the SMXs on the p side of each sensor. The remaining four SMXs on the p side of each sensor remained operational at a single uplink per SMX.
    
    \subsection{High-Energy Test}
    \label{sec:rate_he}
    
    To validate the dynamic range of the sensor with clinically-relevant beams, a final calibration measurement was conducted at the Heidelberg Ion Beam Therapy facility, using a clinical \qty{206.91}{\MeV \per \nucleon} \cion{} beam incident on a \qty{1.0}{\mm} PMMA foil. Again, the fIVI Range Monitoring System was placed at a \ang{40} angle, with the foil centered in its field of view, this time with a \qty{204.6}{\mm} distance between the foil and the front sensor. The entire setup was aligned manually, using the laser system integrated into the experimental cave at HIT, as shown in \figref{setup_he}. Irradiations were completed at primary beam intensities of \qtyrange[range-phrase=~and~]{2e6}{1e7}{\ion\per\s}, for multiple complete synchrotron spills. Both sensors were biased as in previous tests. For this test, all uplinks were again active, with two uplinks per SMX for the front sensor and the remaining uplinks evenly split between the p and n side of the rear sensor. This change in topology from the low-energy test was selected to maximize the available bandwidth for the front sensor, as it was found to experience a significantly higher trigger rate than the rear sensor.
    
    \singleFig[!t]{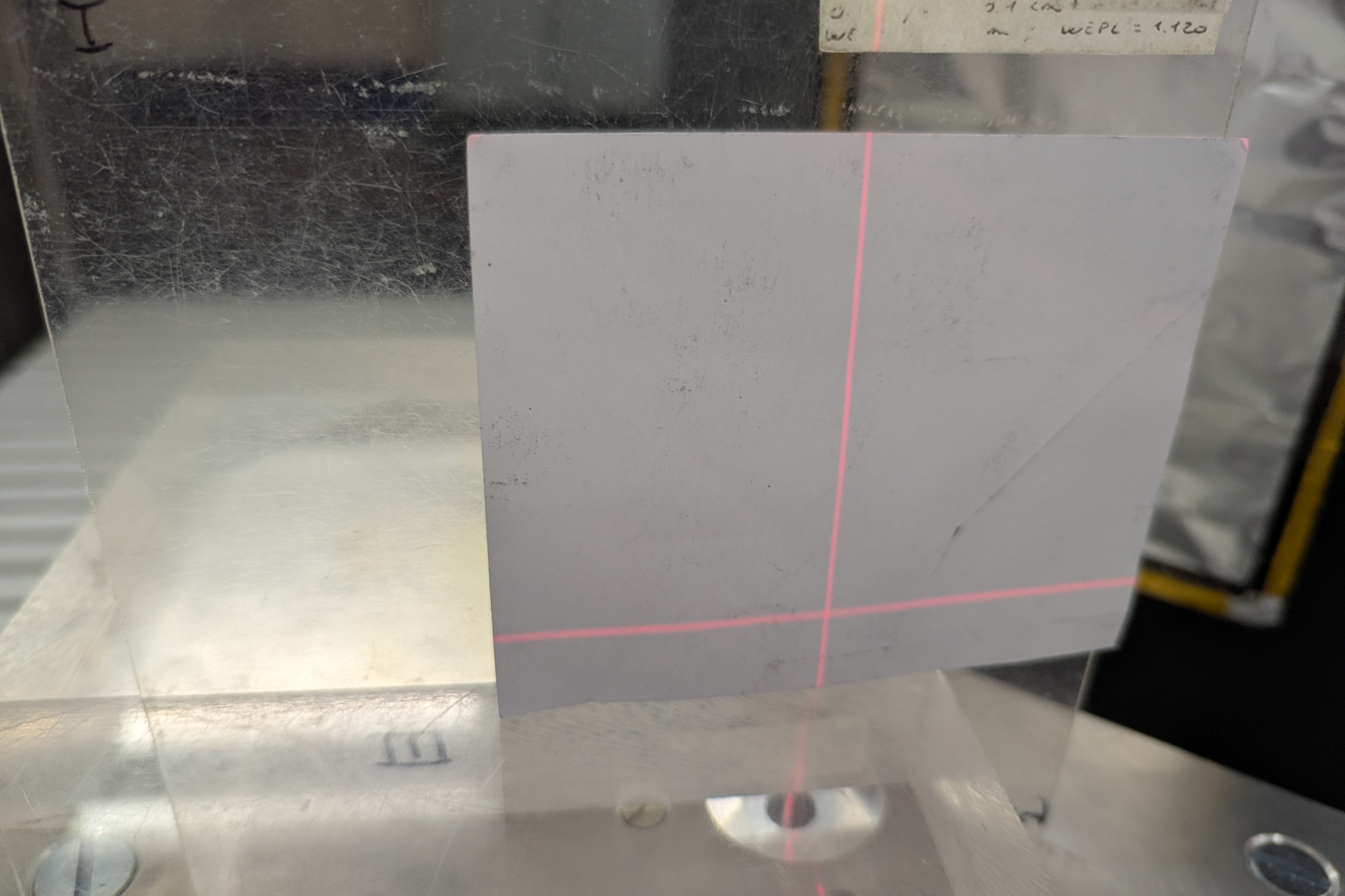}{setup_he}{Detail view of the \qty{1.0}{\mm} PMMA scattering foil used in the high-energy beam test. This foil was installed at HIT in the $Z=0$ plane. The laser alignment system shows the location of the beam isocenter within the foil. The configuration is otherwise similar to the low-energy test, with the tracker placed at a \ang{40} angle from the beam axis, and the front sensor placed \qty{206.4}{\mm} from the isocenter.}{Detail view of the \qty{1.0}{\mm} PMMA scattering foil used in the high-energy beam test.}
    
    \subsection{Data Analysis}
    \label{sec:rate_data_analysis}
    
    All data analysis was completed offline, using code developed specifically for the fIVI Range Monitoring System. Coincident clusters of adjacent strips were found on each sensor side, and coincident clusters from opposite sides of the same sensor were matched to form hits on each module. Coincident hits on both modules were used to form particle tracks, which could be followed back to the target foil plane to reconstruct the interaction position. This analysis code was a subset of the full code designed for clinical range monitoring. Coincidence windows for cluster, hit, and track formation were set to \qty{31.25}{\nano\s}, corresponding to five ticks of the timing clock.
    
    \section{Results}
    \label{sec:rate_results}
    
    \subsection{Low-Energy Rates}
    \label{sec:rate_results_le}
    
    A significant fraction of the total data rate is dedicated to timing information, with the beam-off data rate (with event rates of order \qty{1.0}{\hertz}) being more than \qty{50}{\percent} of the written data rate at even the highest tested beam intensity of \qty{20}{\nano\ampere}. As timing information is maintained independently for each uplink, the exact timing data rate varies depending on the number of active links, at \qty{1.25}{\mega\byte\per\s} per link. 
    
    The fraction of pileup-free events, shown in \figref{le_event_rate}, is reduced at higher beam intensity, ranging between \qty{99.9961}{\percent} and \qty{99.9995}{\percent}; the fraction of events affected by pileup is negligible at all intensities. The lack of a significant difference between the single-link-per-SMX and all-link configurations at similar beam currents of \qty{13.5}{\nano\ampere} and \qty{15}{\nano\ampere} indicates that the observed pileup is primarily due to occasional secondary events occurring before reset was complete, rather than due to a bandwidth limitation in the readout system.
    
    \doubleFig{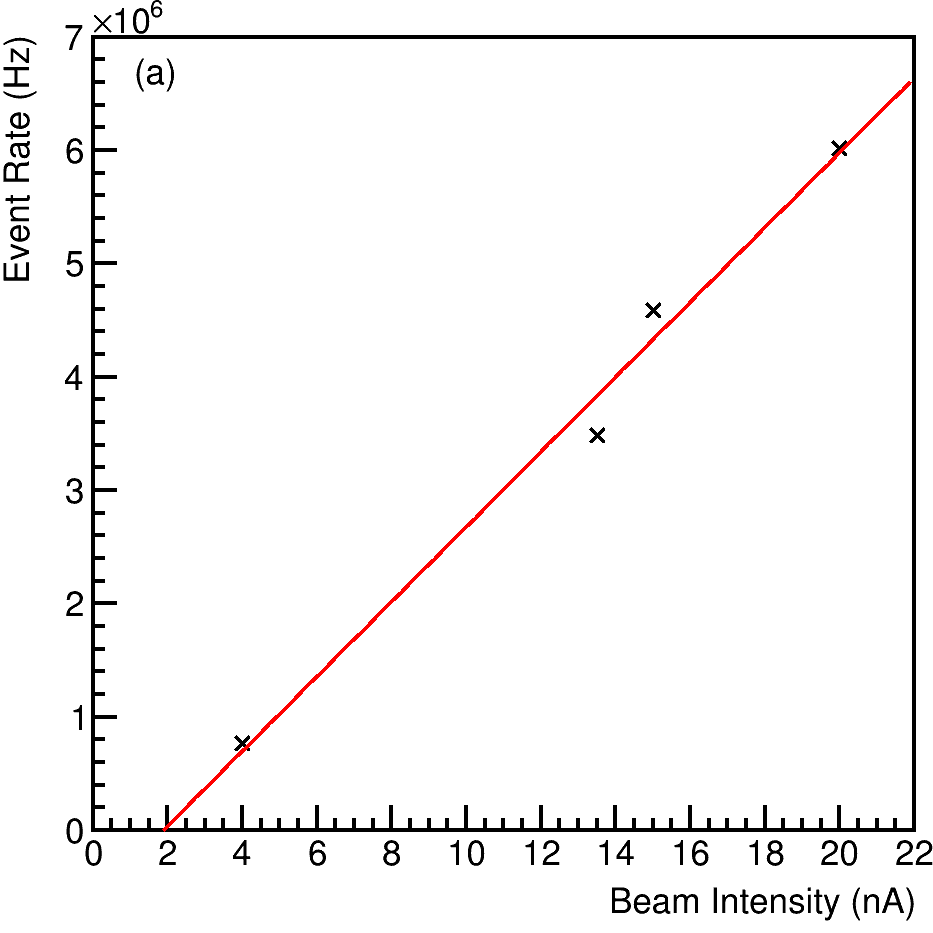}{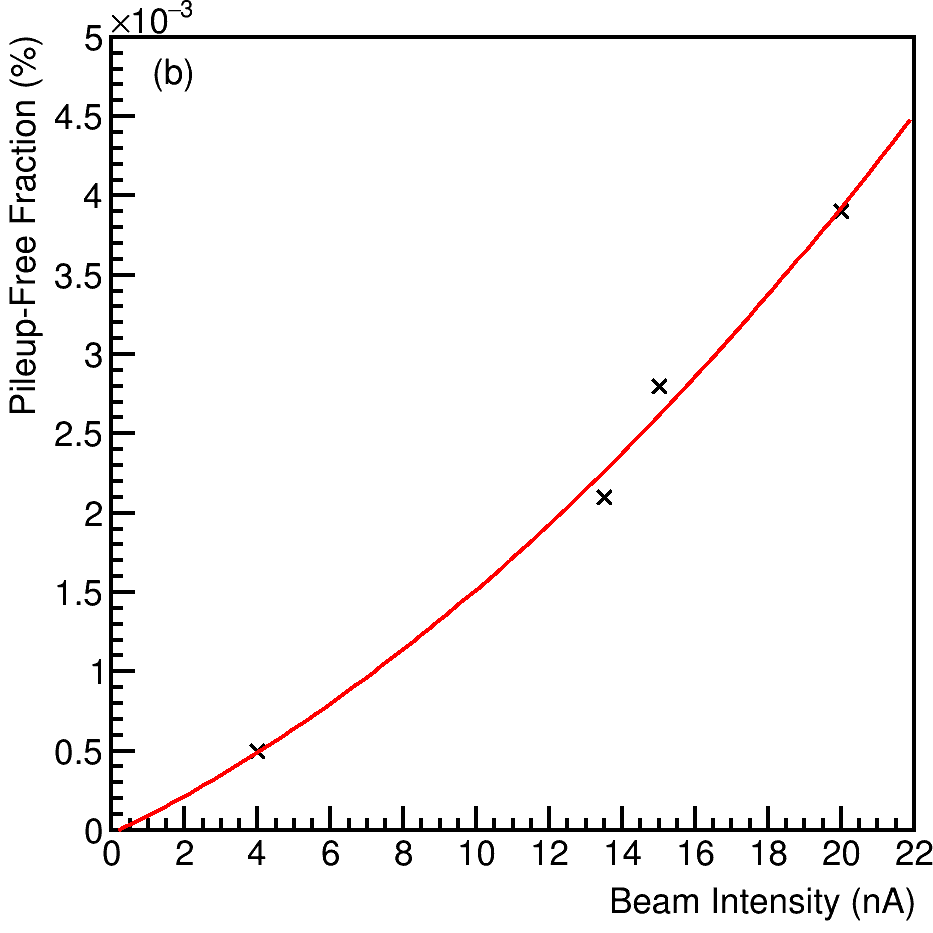}{le_event_rate}{Trigger and pileup rates in the low energy tests. (a) Trigger event rate in the low-energy tests as a function of beam current. The linear trend indicates that the observed event rate is far from the maximum rate which the system is capable of accepting. (b) Fraction of events affected by pileup, as a function of beam current. The quadratic relationship is expected for events with a Poissonian distribution in time.}{Trigger and pileup rates in the low energy tests.}
    
    When controlling for the timing overhead, the data rate is directly correlated to the number of trigger events accumulated, indicating that at the data rates examined, there is no significant change in the number of TS\_MSB frames generated per Hit frame. The rate of cluster formation on the front sensor is \num{2.32 \pm 0.01} times that of the rear sensor, while the ratio in solid angle coverage between the two sensors is only approximately \num{1.33}. However, as the distribution of events is highly nonuniform, this deviation is not unexpected.
    
    The rate of cluster formation on the p side of the sensor, with the angled strips, is consistently \qty{2}{\percent} greater than the rate on the n side for the front sensor, and \qty{5}{\percent} greater for the rear sensor. The reason for this difference is not immediately clear. One possible explanation is systematic differences in the achievable minimum thresholds and noise levels between the p and n sides of the sensors. Due to the significantly longer conductor length associated with the interconnect, the Z-strips exhibit a significantly higher noise level than either axial or angled strips which do not have an interconnect \autocite{rodriguez_rodriguez_advancements_2025}. This higher noise level may result in both a greater number of false events due to noise, or a greater sensitivity to events with low deposited energy. The number of Z-strips is doubled on the rear sensor as compared to the front sensor, due to the sensor length being doubled; this relationship corresponds to the approximate doubling in the rate of excess cluster formation on the p side.
    
    The number of hits formed on each sensor scales linearly with the number of clusters formed on the n side of that sensor, with losses of \qty{2}{\percent} or less indicating a high success rate in matching clusters between the p and n sides of the sensor. As the number of hits does not exceed the number of clusters, there are few coincidence windows in which ghost hits may form. The ratio of hits formed on the front and rear sensor is \num{2.44 \pm 0.03}, similar to the ratio of cluster formation. The average hit rate on the front sensor at the \qty{20}{\nano\ampere} irradiation was \qty{1.3}{\mega\hertz}, matching the target event rate.
    
    The rate of track formation is similar to the rate of hit formation on the rear sensor, reflecting the field-of-view limiting effect of this sensor. The typical ratio of track formation to rear sensor hit formation is \qty{88.8 \pm 1.7}{\percent}. A ratio of approximately \qty{100}{\percent} is indicative of most coincidence windows between the front and rear sensor containing only a single hit on each layer. In this configuration, not all particles travelling from the origin to the upper and lower edges of the rear sensor will interact with the front sensor, leading to those hits not being expected to contribute to track formation. When the hits from these sections of the rear sensor are neglected, the track formation rate rises to just over \qty{100}{\percent}. A comparison between the formation rates of clusters, hits, and tracks is presented in \figref{rates_le}.
    
    \singleFig[!b]{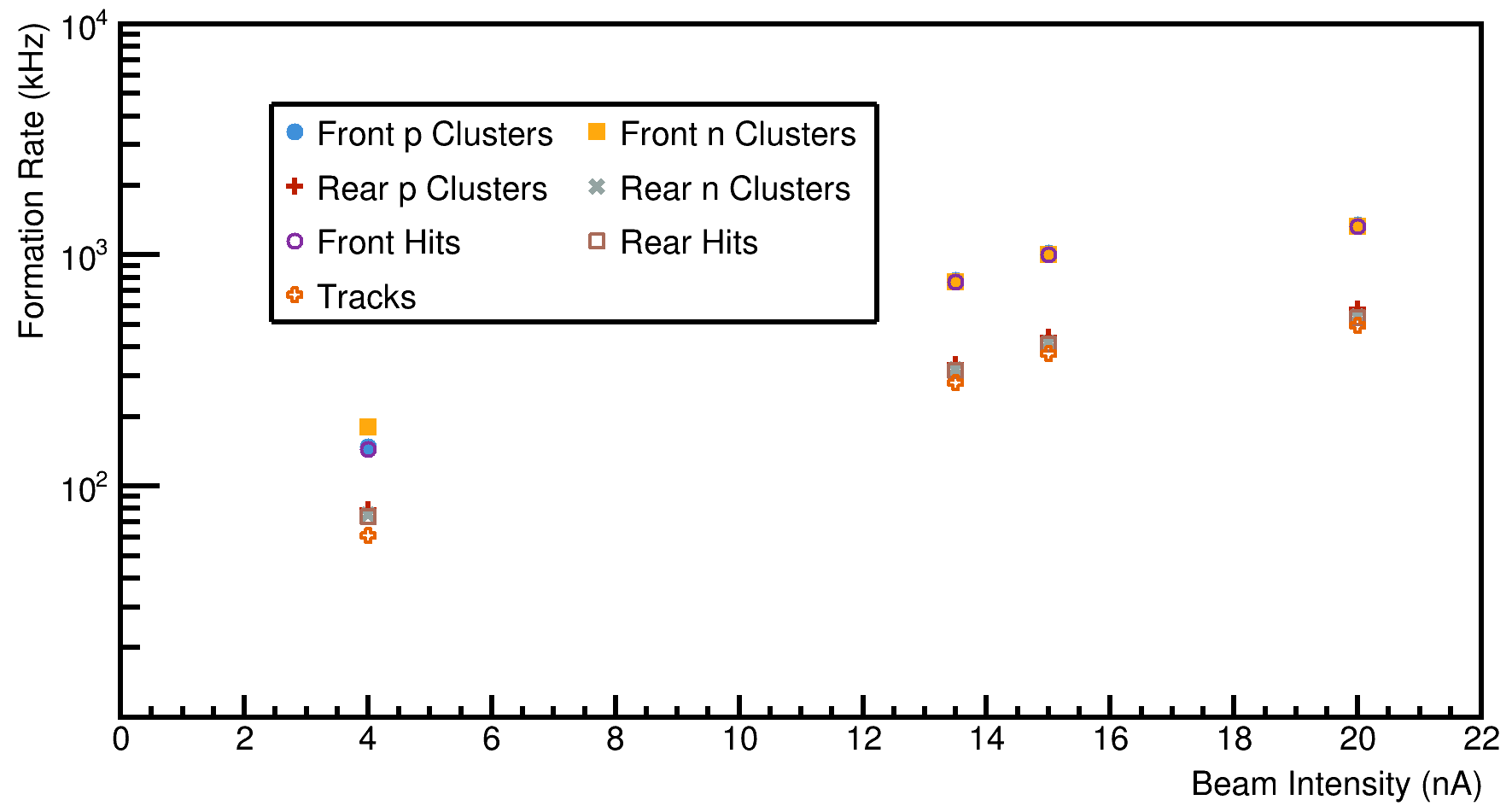}{rates_le}{Comparison between formation rates in the low-energy test as a function of beam current. Clusters on each sensor side, hits on each sensor, and tracks are plotted in separate series. Each data set exhibits a linear trend. The close agreement between different variables shows the high efficiency of the reconstruction process, indicating that very few events are recorded which do not correspond to a trackable particle.}{Comparison between formation rates in the low-energy test as a function of beam current.}
    
    The distributions of cluster sizes are similar between both sides of both sensors. Overall, cluster sizes of 1 and 2 are most common, with the proportion of larger clusters decreasing exponentially.
    
    \subsection{High-Energy Rates}
    \label{sec:rate_results_he}
    
    As the clock frequency for both configurations is the same, there is no expected or observed difference in the data rate dedicated to timing information, remaining at \qty{1.25}{\mega\byte\per\s} per uplink. Due to beam intensity limits in the HIT standard beam library, the absolute intensity for these irradiations is significantly lower than the low-energy tests, which were intended to model the yield from a beam path length orders of magnitude longer than the single millimeter thickness of the target foil. Correspondingly, the data rate associated with trigger events is negligibly low, on the order of hundreds of \unit{\kilo\byte\per\s}. As with the low-energy beams, the fraction of events lost to pileup is negligibly low, with fewer than 10 lost events in total, between the two irradiations.
    
    The rate of cluster formation is again greater on the p side of the sensor than the n side, although by a significantly larger fraction than for the low-energy beams. This discrepancy is attributed to a marked increase in the proportion of clusters produced with the lowest possible energy. These low-energy clusters are produced significantly more often on both the p and n sides of the sensor, as compared to the low-energy beam tests. Particularly on the p side of the front sensor, the number of low-energy clusters is greatly enhanced, with a clear variation visible between SMXs. This pattern indicates an issue with the threshold on the lowest-energy discriminator of the ADC being slightly too low, as these thresholds naturally vary slightly between individual ASICs. The same higher noise level on the p side owing to the Z-strips is also believed to contribute, although the effect size is smaller than the threshold configuration issue. The ratio of cluster formation between the front and rear sensors is approximately \num{2.426 \pm 0.003} when considering n side clusters only. As the p side of the front sensor has a particular enhancement in low-energy clusters, the ratio between p sides is much larger, at \num{5.48 \pm 0.01}.
    
    The rate of hit formation again follows the n side distribution of clusters, although the rate is suppressed due to the additional noise-based clusters at low deposited energies. On the front \qtyproduct[product-units=power]{60x60}{\mm} sensor, approximately \qty{89.5}{\percent} of clusters match with the opposite side to form a hit, while on the rear sensor, approximately \qty{96.6}{\percent} of clusters form hits. The ratio between the front and rear sensors is consistently \num{2.248 \pm 0.002}. This ratio is slightly lower than for the low-energy beam test; however, this difference is not unexpected due to the few centimeters difference in the distance between the front sensor and the scattering foil. The overall hit rate at the highest beam intensity of \qty{e7}{\ion\per\s} is approximately \qty{7}{\kilo\hertz}.
    
    The rate of track formation is significantly suppressed as compared to the hit formation rate, with only \qty{48.3 \pm 0.8}{\percent} of hits on the rear sensor forming tracks. This suppression, shown in \figref{rates_he}, is partially attributable to the enhancement in the number of clusters formed at low deposited energies, with only a negligible fraction of these clusters continuing all the way to the track formation process. However, this phenomenon alone is insufficient to explain the low fraction of rear sensor hits contributing to track formation. Another possible explanation is the interaction of scattered beam particles with the tracker enclosure, with those particles scattering a second time, or producing fragments, which then interact with the rear sensor only.
    
    \singleFig{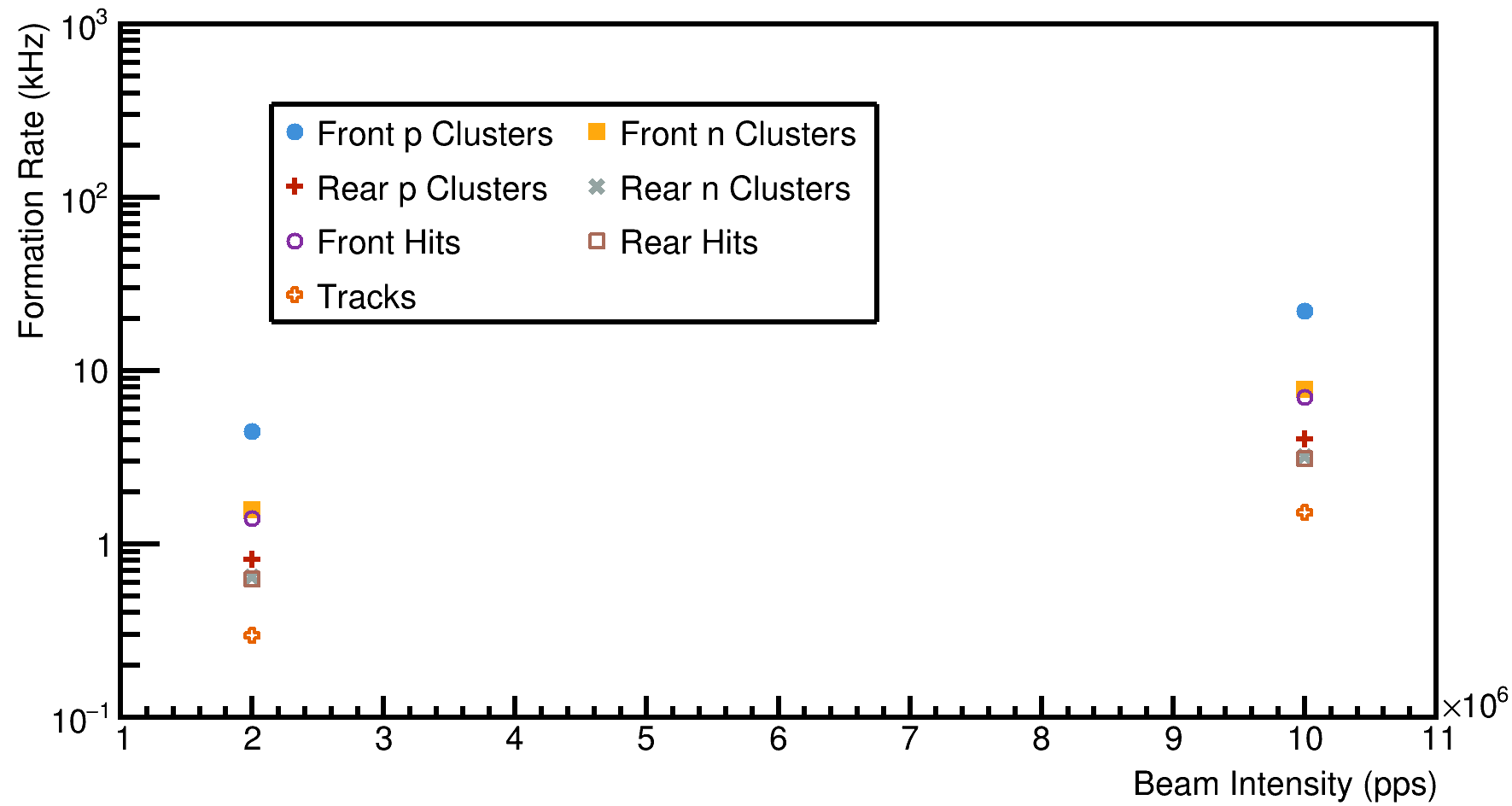}{rates_he}{Comparison between formation rates in the clinical beam test as a function of beam current. Clusters on each sensor side, hits on each sensor, and tracks are plotted in separate series. The elevated rate of cluster formation on the p side of both sensors is attributed to an incorrectly-set threshold, while the suppressed rate of track formation relative to hit formation is attributed to interactions on the rear sensor which do not directly correspond to a particle originating from the scattering foil.}{Comparison between formation rates in the clinical beam test as a function of beam current.}
    
    \subsection{Spatial Distribution}
    \label{sec:rate_spatial}
    
    For the low-energy test, the overall distribution of interactions on the sensor is shown in \figref{rutherford}. This distribution is dominated by Rutherford scattering, which results in a significantly enhanced signal at smaller angles. The bulk of the spatial dependence of event rate is explained by the Rutherford distribution, although an enhancement at large angles is observable. This deviation from pure Rutherford scattering is attributed in part to the thick target, through the influence of multiple scattering; this phenomenon is confirmed by the Geant4 simulation. However, a further small enhancement at large angles is observed in the experimental data; this increase is believed to be from secondary scattering at the edges of the beam pipe exit window, which is not modelled in the simulation. As the angular variation is greater for the front sensor than the rear, due to the smaller separation from the scattering foil, this phenomenon is more prominent when considering interactions on the front sensor only.
    
    \singleFig[p]{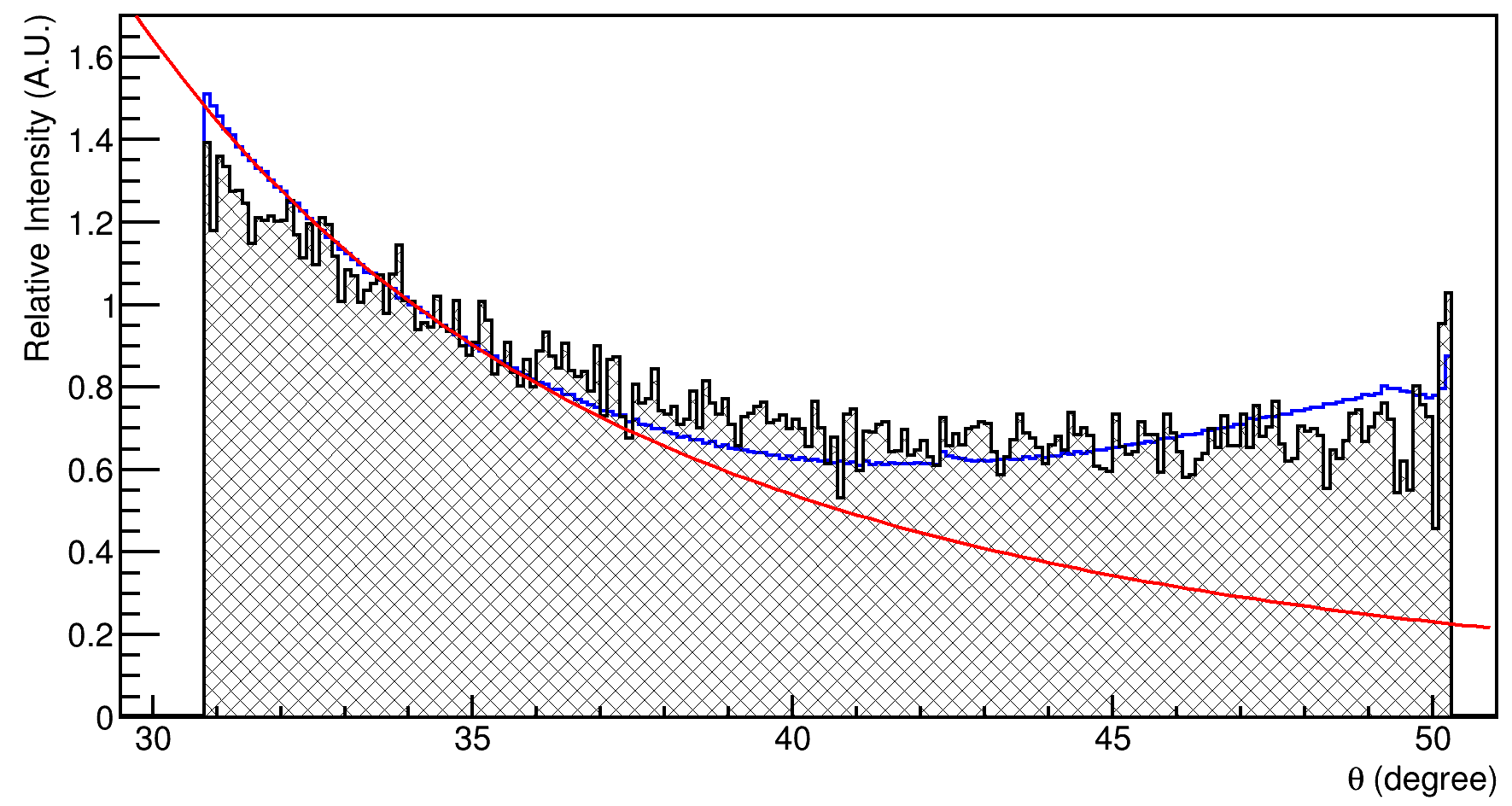}{rutherford}{Angular distribution of events (blue) on the front sensor during low-energy beam test. Data has been normalized to account for the variation in sampling region as a function of angle. The red line represents the expected distribution from pure Rutherford scattering, and is scaled to match the experimental distribution at small angles. The enhancement at larger angles is attributed to the target thickness, as is reproduced by the Geant4 simulation (black, hatched).}{Angular distribution of events on the front sensor during low-energy beam test.}
    
    As the spatial distribution of interactions is highly nonuniform, the distribution of event rates between individual SMX ASICs, shown in \figref{asic_rate}, is also highly nonuniform. The highest rates are visible on the front sensor, from ASICs which read out the edge at the smallest angle from the beam axis. Maximum event rates are similar on both the p side and n side of each sensor, with the most active ASIC in each case recording a trigger rate of \qty{480}{\kilo\hertz} at \qty{20}{\nano\ampere}. However, the most active links consistently appear on the p side, where some ASICs are limited to a single uplink. The reduction in trigger rate on the rear sensor is more prominant than the reduction in cluster formation rate, with the most active ASICs on the rear recording a trigger rate of \qty{150}{\kilo\hertz} due to the larger off-axis angle. Similar trends exist for the high-energy beam test; however, the elevated noise level on the p side of the sensor makes absolute ASIC-level comparisons more difficult.
    
    \doubleFig[p]{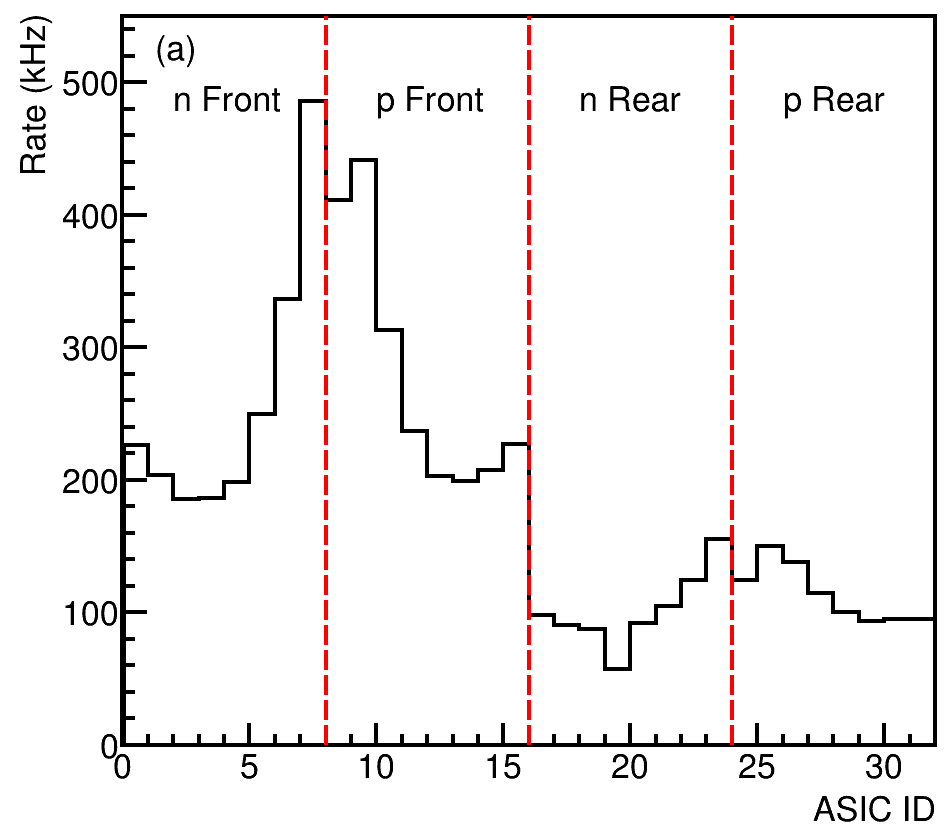}{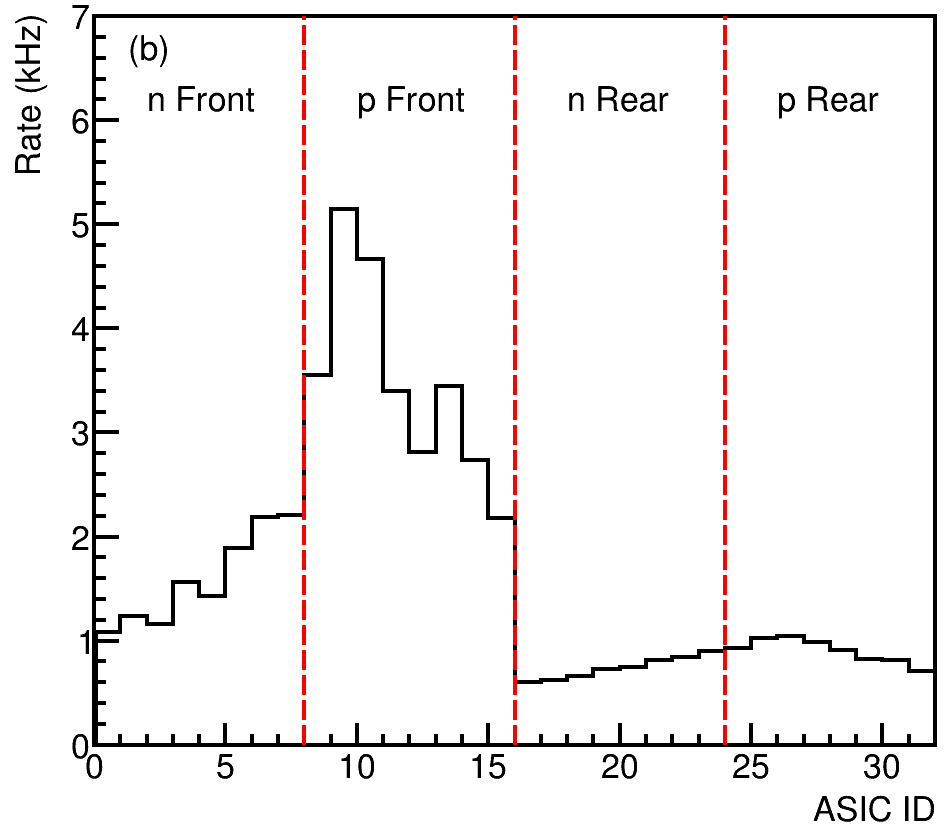}{asic_rate}{Trigger event rates for each ASIC for (a) low-energy test and (b) clinical beam test. Higher ASIC IDs on the n side of the sensor correspond to smaller off-axis angle $\theta$, while higher ASIC IDs on the p side of the sensor correspond to larger off-axis angle $\theta$. Peak event rates on the n side of the sensor are typically slightly higher than the p side, as the angled strips on the p side result in each strip covering a wider range of values of $\theta$, while the event rate is heavily $\theta$-dependent. Event rates on the p sides of each sensor are artificially elevated for the clinical beam test, due to the increased noise level in this environment.}{Trigger event rates for each ASIC.}
    
    The cluster size also tends to vary as a function of position on the sensor, as shown in \figref{cluster_size}. This position dependence is believed to be attributable to variation in the typical angle at which a particle passes through the sensor. Along the horizontal axis, clusters closer to the center are more likely to exhibit a cluster size of 1, where normal or nearly-normal incidence is more frequent. In contrast, clusters closer to the edge of the sensor are more likely to exhibit a cluster size of 2, due to the larger average angle of incidence. Although cluster sizes larger than 2 are significantly less common, these larger cluster sizes are again more prevalent towards the outer edges of the sensor. Along the vertical axis, the distribution of cluster sizes is much more uniform across the entire sensor, as the vertically-oriented segments are less sensitive to variations in vertical position or angle. The same patterns hold for clusters appearing on both the p and n side of the sensor, as the orientation of segments is similar. Again, these patterns are more prominent for the front sensor than for the rear. At clinical beam energies, cluster size 1 events are more heavily represented, due to the threshold differences discussed in \secref{sec:rate_results_he}.
    
    \quadFig[p]{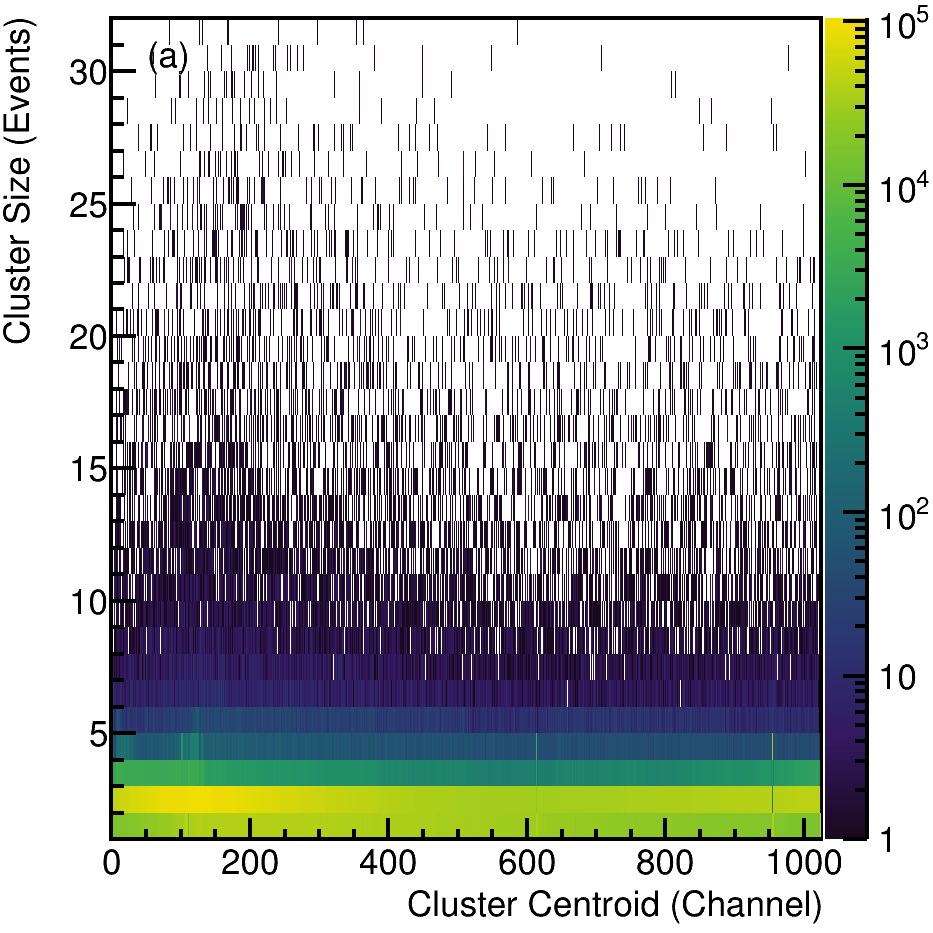}{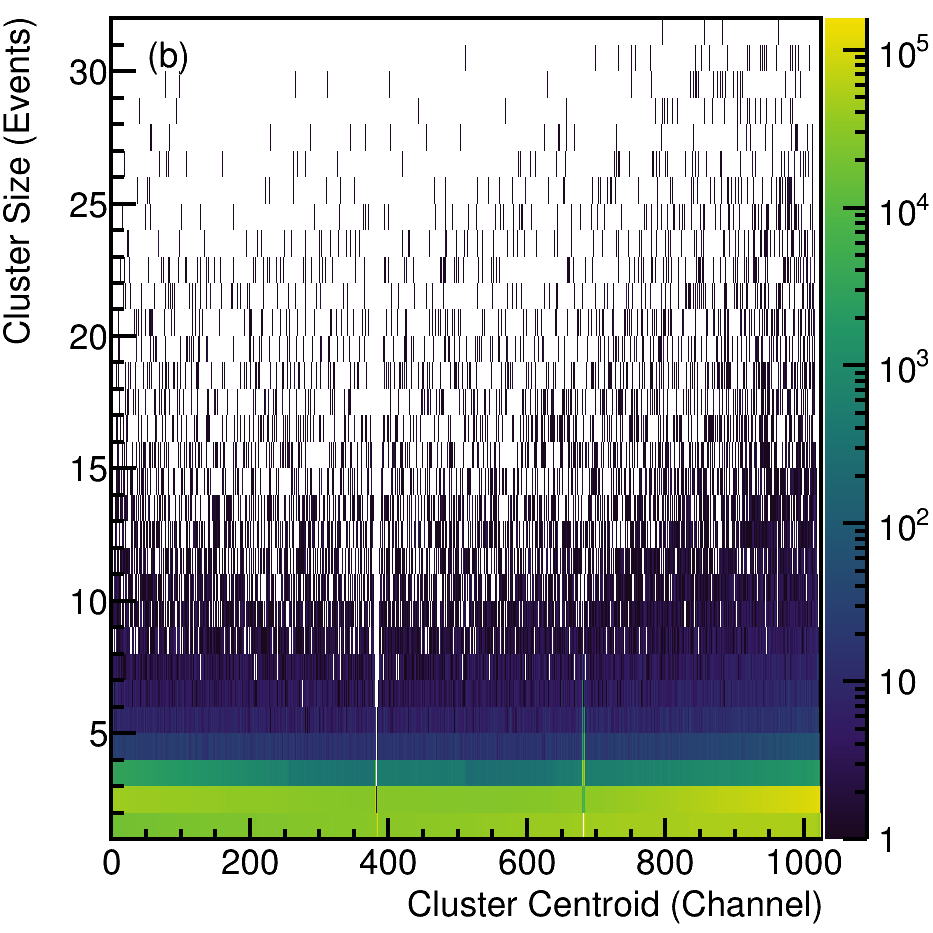}{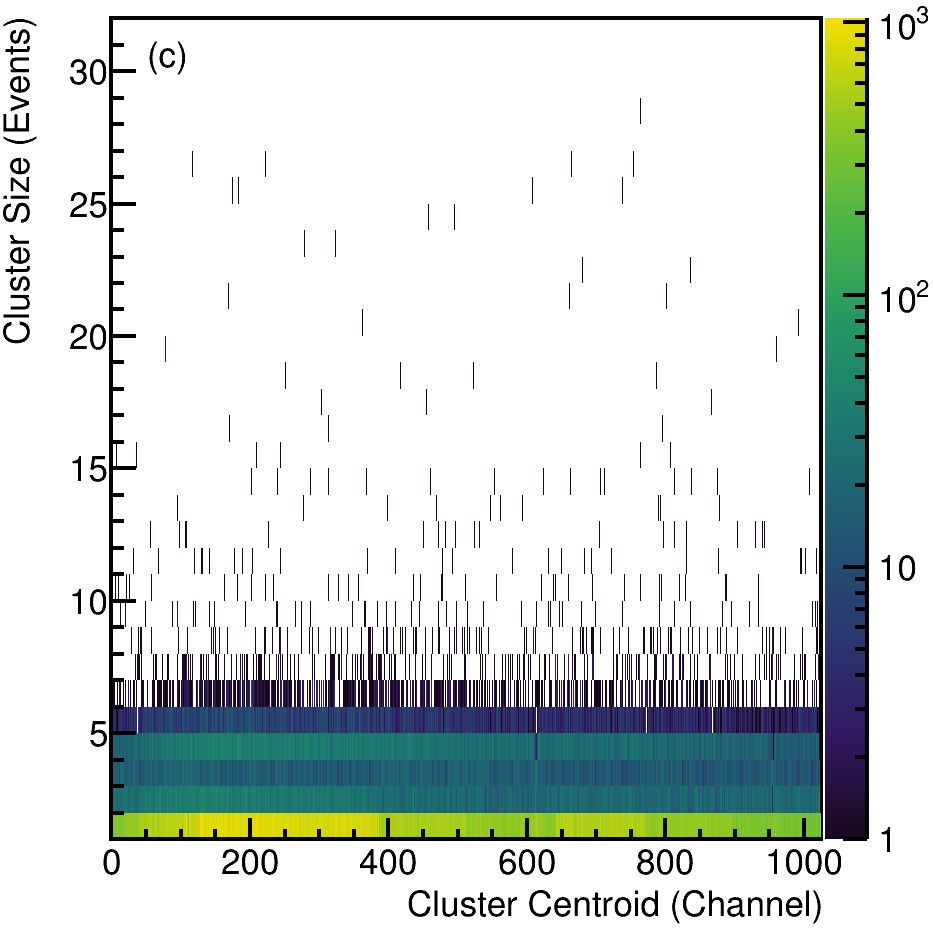}{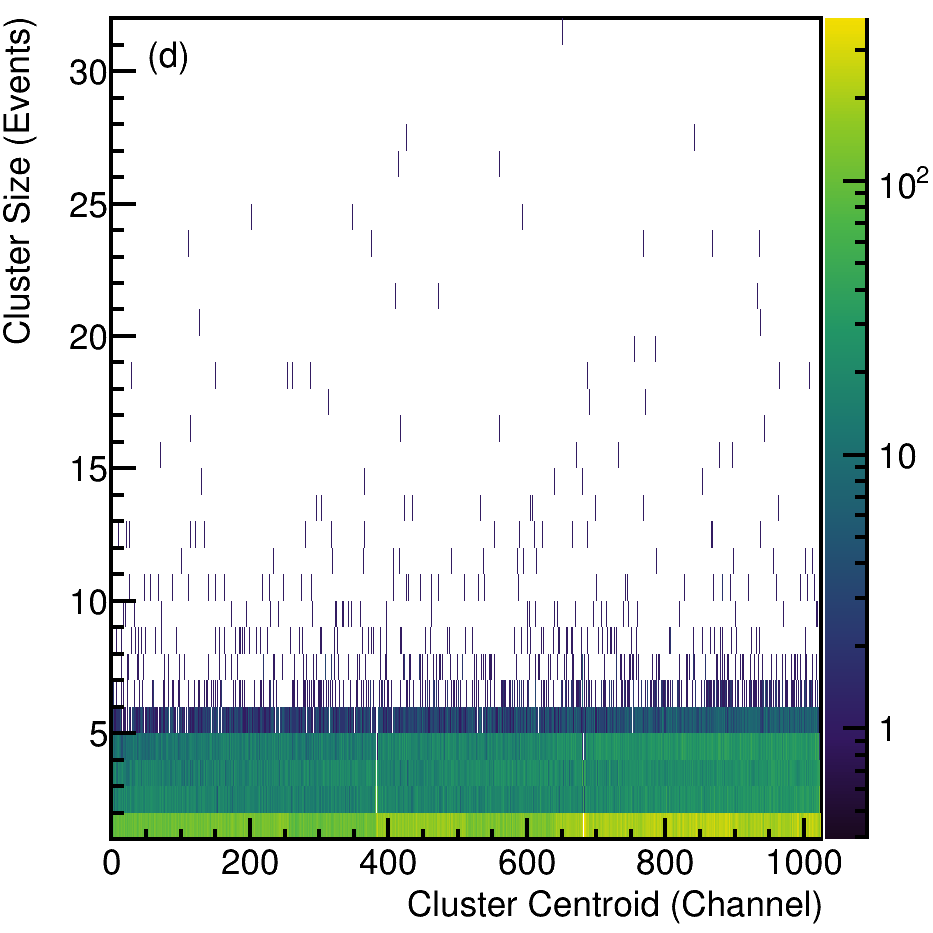}{cluster_size}{Cluster size distributions on the front sensor for each test as a function of effective channel number. Artifacts are visible at certain channels at the boundary between ASICs, where the system is more susceptible to noise. Some noisy channels lead to an inflation in average cluster size, while other noisy channels have been masked during analysis, leading to apparent gaps in the data. (a) Distribution from the low-energy test on the p side of the sensor, with angled strips. The Z-strips appear at lower channel numbers, which also correspond to a smaller off-axis angle $\theta$. (b) Distribution from the low-energy test on the n side of the sensor, with axial strips. Higher channel numbers correspond to a smaller off-axis angle $\theta$. (c) Distribution from high-energy test on the p side of the sensor. (d) Distribution from high-energy test on the n side of the sensor.}{Cluster size distributions on the front sensor.}
    
    \subsection{Beamspot Reconstruction}
    \label{sec:rate_beamspot}
    
    In the low-energy tests, the beamspot is consistently reconstructed close to the expected position at the origin of the coordinate system, with positioning errors of less than \qty{2.5}{\mm} from the nominal center point. This deviation is consistent with the position accuracy of beam delivery to the setup, and is not believed to indicate a significant inaccuracy in tracking. A representative beamspot is shown in \figref{beamspot_le_a}. There is a notable asymmetry in the shape of the beamspot along the horizontal axis, with an enhancement of the signal in the direction of the positive axis, away from the position of the tracker. This asymmetry is believed to be due to secondary scattering of fragments from the edge of the window in the helium-flooded beam pipe, the same as the source of the enhancement at large angles observed in the hit patterns. To reduce the effect of this scattering, a filter was applied at the level of track formation. Under this filter, two coincident hits are only allowed to form a track if the difference in polar angle between the two hits is \ang{1.0} or less, and the difference in azimuthal angle is \ang{2.0} or less.
    
    \doubleFig[!b]{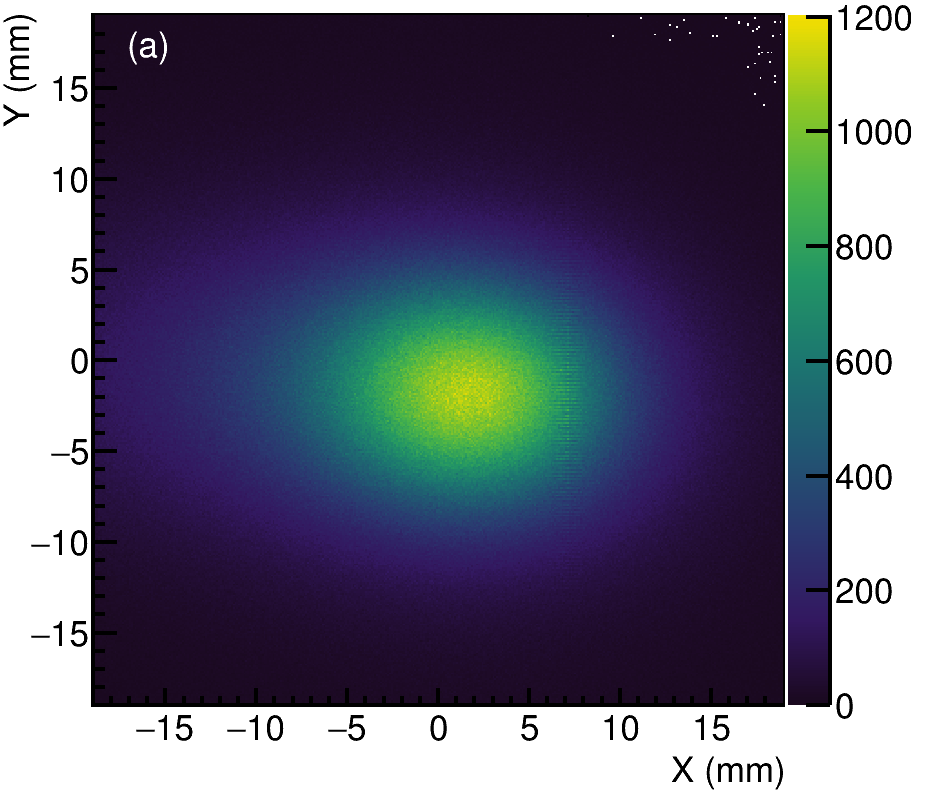}{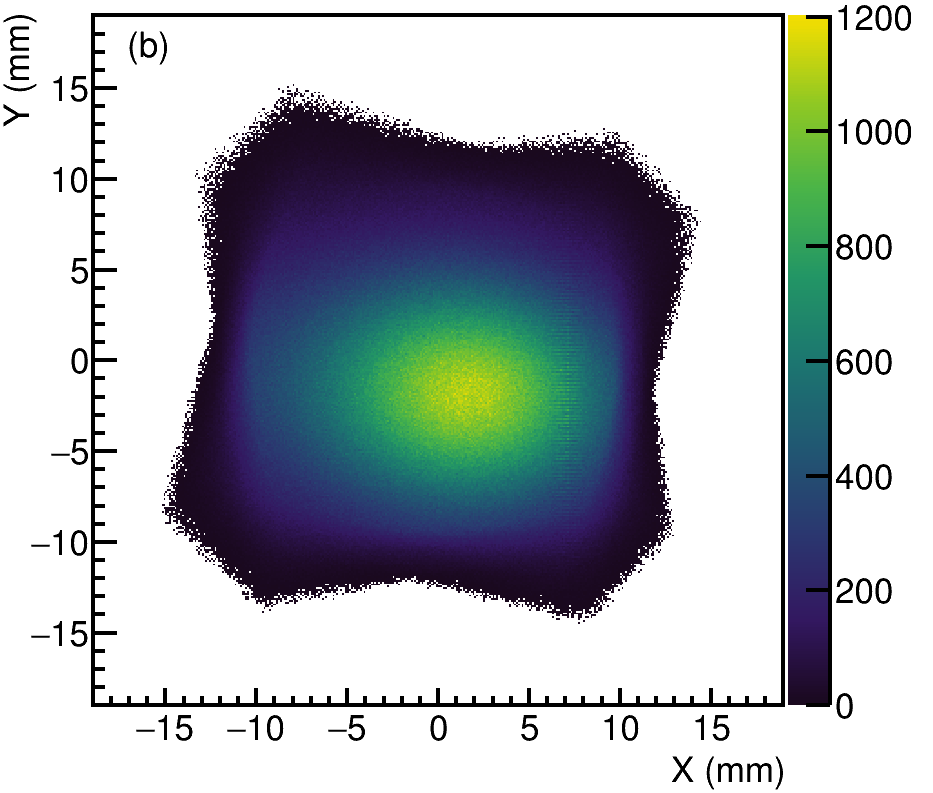}{beamspot_le}{Reconstructed beamspot images from low-energy test. (a) Reconstructed beamspot without the angle filter applied. The beamspot is approximately Gaussian in shape, with a horizontal standard deviation of \qty{7.9411 \pm 0.0017}{\mm}, and a vertical standard deviation of \qty{5.387 \pm 0.0010}{\mm}. (b) Reconstructed beamspot with the angle filter applied. The rotational symmetry of the beamspot is enhanced, with a horizontal standard deviation of \qty{5.7871 \pm 0.0013}{\mm} and a vertical standard deviation of \qty{4.5752 \pm 0.0010}{\mm}. In both cases, the beamspot is offset horizontally from the expected position by less than \qty{0.5}{\mm}, and vertically by approximately \qty{1.5}{\mm}.}{Reconstructed beamspot images from low-energy test.}
    
    The size of the filtered reconstructed beamspot shown in \figref{beamspot_le_b}, is significantly larger than the \qty{3.0}{\mm} focal size of the incident proton beam. However, this increased beamspot size is not entirely unexpected. Firstly, due to the use of the scattering foil as an exit window, beam focusing to the experimental setup was completed approximately \qty{30}{\cm} in front of the foil, allowing for some defocusing to take place. Secondly, due to the low energy of the Rutherford-scattered protons, significant further scattering is possible in the first detector layer, which will increase the apparent beamspot size. Some of the effect of this scattering in the first detector layer is also mitigated by the same filter which reduced the asymmetry due to beamline scattering.
    
    In the high-energy tests, the reconstructed beamspot, shown in \figref{beamspot_he}, is again located close to the expected position at the isocenter, with a horizontal deviation of \qty{2.555 \pm 0.019}{\mm}, and a vertical deviation of \qty{1.211 \pm 0.018}{\mm}. This deviation is within the positioning uncertainty of the setup using the laser alignment system. This beamspot exhibits the expected symmetry, as there were no beam pipe elements in this setup from which secondary scattering could occur. The full width at half maximum (FWHM) values of the reconstructed beam are similar to the measured beamspot size by the HIT beam monitoring system, which slightly exceeded the nominal beamspot size of \qty{5.0}{\mm} FWHM.
    
    \singleFigNarrow[!t]{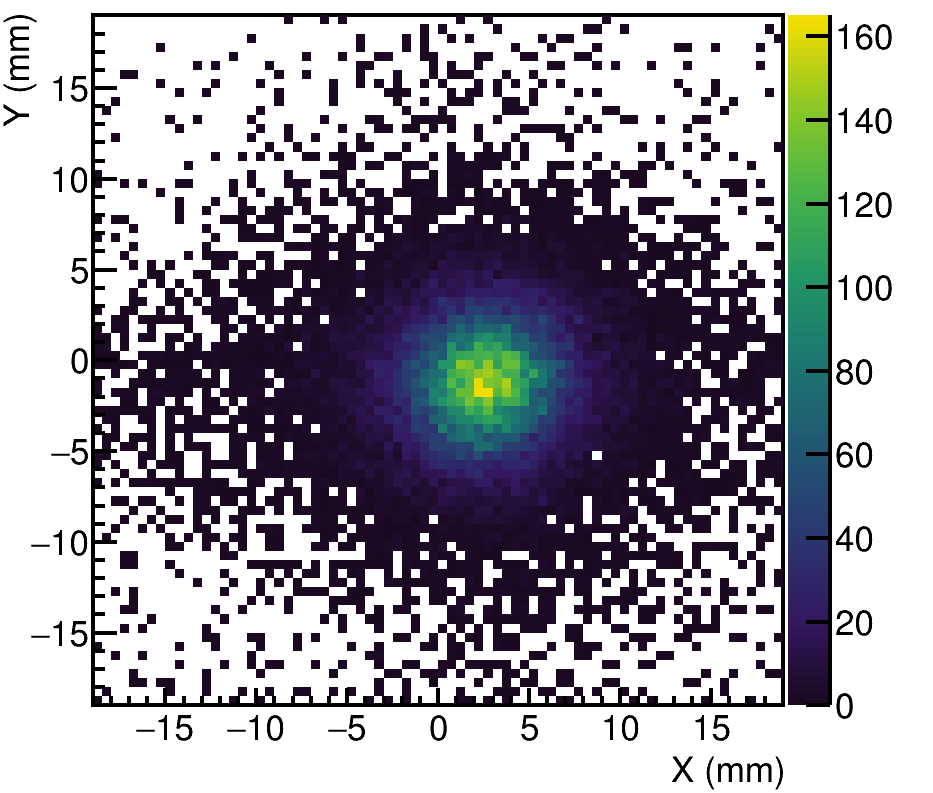}{beamspot_he}{Reconstructed beamspot image from clinical beam test. The beamspot is again approximately Gaussian in shape, with a horizontal standard deviation of \qty{3.119 \pm 0.018}{\mm} and a vertical standard deviation of \qty{2.890 \pm 0.015}{\mm}. These values correspond to a full width at half maximum of \qty{7.35}{\mm} horizontally, and \qty{6.81}{\mm} vertically.}{Reconstructed beamspot image from clinical beam test.}
    
    \section{Discussion}
    \label{sec:rate_discussion}
    
    The event rate measured in the low-energy tests meets and exceeds the calculated \qty{1.2}{\mega\hertz} event rates which are expected to occur during irradiation of a human-scale target with a clinical \cion{} beam. While this rate is not directly comparable to those measured in the high-energy beam test, the discrepancy is primarily due to the difference in target thickness between the beam test and a clinically-relevant irradiation. This difference in thickness has a nontrivial impact on the total measured event rate, as a thicker target provides a greater opportunity for the beam to fragment, but also introduces more material in which fragments with lower energies may slow and stop before reaching the tracker.
    
    The similar tracking and reconstruction performance displayed between the low-energy and high-energy beam tests highlights the suitability of this sensor and readout system to operate in a large range of energy regimes. While the noise level is significantly higher for the high-energy beam test data, this challenge is expected to be solvable through small adjustments to the threshold level on each SMX. While such a calibration process is time-consuming, it would only need to be performed once for a given energy regime \autocite{saini_test_2023}. After this calibration adjustment, the progress rates through the different phases of data analysis and reconstruction are expected to be more in line with those of the low-energy data. Indeed, simply eliminating the trigger events with the lowest possible energy is already sufficient to produce cluster formation results significantly closer to the excellent reconstruction efficiency observed in the low-energy tests.
    
    The reduced track formation efficiency in the high energy tests is not believed to be characteristic of a performance issue, as the efficiency remains similar regardless of changes to the primary beam intensity. It is believed that this reduced efficiency is instead related to true interactions on the rear sensor which have no counterpart on the front sensor, from particles which have not originated at the scattering foil. Such particles could be produced by the interaction of primary ions which undergo small-angle scattering at the target foil, and interact with the tracker enclosure, undergoing fragmentation or multiple scattering to interact with the rear sensor only. This phenomenon would not be expected to occur in the low-energy tests, due to the beam pipe blocking the majority of small-angle scattered particles before they could reach other elements of the setup. During a clinical irradiation, these effects are expected to be significantly reduced, as the primary treatment beam and many forward-focused fragments stop in the patient, reducing the background with which the tracker must contend.
    
    All tested configurations remained well below the theoretical rate limits of the sensor and readout system. Under the tested conditions of an \qty{80}{\mega\hertz} uplink clock, each uplink is expected to provide a bandwidth of \qty{5.33}{\mega \frame \per\s}. At high rates, one TS\_MSB frame is expected to be generated for every 7.5 trigger frames, leading to an effective bandwidth of \qty{4.71}{\mega \Hit \per\s}, for trigger events over the entire 128-channel ASIC \autocite{kasinski_characterization_2018}. However, as interactions of the primary beam are typically not uniformly distributed in time \autocite{peters_spill_2008}, it is recommended to run the system at only \qty{67}{\percent} average utilization, suggesting an average data rate of \qty{3.2}{\mega \Hit \per\s} should be the maximum targeted. This rate is a factor of six above the maximum experienced in the low-energy test, indicating that there is still significant headroom in the design. Even with the expected increase in the fraction of events lost to pileup, the fraction of lost events would still not affect a significant fraction of the data.
    
    Further increases in rate could be achieved through either an increase in the uplink clock rate, or an increase in the number of uplinks per SMX. Each additional uplink, up to a total of five, would provide a linear increase in the data rate, as each link must communicate its own timing data using TS\_MSB frames. The uplink clock rate is also capable of being doubled to \qty{160}{\mega\hertz}, with a corresponding increase in timestamp precision and overall bandwidth. Maximizing both of these variables would result in a total trigger event bandwidth of \qty{32}{\mega \Hit \per\s} for each ASIC, at the same \qty{67}{\percent} link utilization target \autocite{kasinski_characterization_2018}. However, when fully utilizing this maximum bandwidth, the fraction of missed events is expected to increase to over \qty{6}{\percent}, which is projected to have a significant impact on the track formation efficiency, due to the multiple coincidences required at various levels of the reconstruction process. Fortunately, these effects are not expected to become significant until beam intensities are more than an order of magnitude beyond that of conventional clinical irradiation.
    
    Due to hardware limitations of the current prototype setup, any increase in total link count or signalling rate would require additional data processing board hardware. However, in all tested configurations, the interaction rate on the two sensors consistently differed by a factor of at least two. As the data readout rate is correlated to the interaction rate, the bandwidth required by the rear sensor is therefore expected to be less than half of that required by the front sensor in this tracker configuration. This finding suggests a possible optimization in the readout system for future tests, or for a purpose-built clinical tracking system: using two \qty{80}{\mega\hertz} uplinks per SMX for all ASICs on the front sensor, and only one uplink per SMX for all ASICs on the rear sensor. Such a configuration would approximately match the available bandwidth for each sensor to the expected data rate, and would require only 48 total uplinks, as opposed to the 56 used in this work. Reducing the total number of uplinks will have a significant impact on the total data rate entering the DAQ PC, optimization of which may be beneficial to online analysis.
    
    Monitoring of FLASH irradiation poses additional challenges in terms of the high particle flux and corresponding data rate. Typical definitions suggest that the FLASH effect occurs at dose rates above \qty{40}{\gray\per\s}, which corresponds to a minimum beam current of \qty{2e9}{\ion \per\s} to irradiate a \qtyproduct[product-units=power]{1.0x1.0}{\cm} cross-sectional area, with higher beam intensities required for larger target volumes or higher dose rates \autocite{tinganelli_ultra-high_2022}. This minimal FLASH beam intensity is expected to be within the capabilities of the prototype system, as the observed cluster sizes are low, which would allow sufficient headroom in the rate capabilities without increasing either the clock frequency or the uplink count. However, the FLASH performance of the prototype system is still expected to be quite sensitive to uniformity in beam intensity, which has so far been found to be less consistent at FLASH intensities than for conventional irradiation \autocite{schoemers_christian_beam_2023, tinganelli_ultra-high_2022}. Operating at higher event rates may also lead to an increase in the rate of formation of ghost hits, which may force further efficiency losses by requiring tighter coincidence windows. Although initial results are promising, further experiments are required to determine the degree to which these complicating factors may affect the ability of this system to perform range monitoring for FLASH irradiation, and any corresponding limits on the dose rate or target volume which may be monitored.
    
    For higher-rate or larger-volume FLASH irradiation, it may also be beneficial to explore the use of the full bandwidth available to the SMX. As a maximum of five uplinks are possible from each SMX, an optimal configuration for FLASH might use five uplinks per ASIC on the front sensor, and two or three uplinks per ASIC on the rear, for a total of 112 or 128 uplinks for the complete tracker. As only two uplinks per ASIC are electronically connected to the current readout PCB, such a setup would at minimum require an upgraded front sensor, in addition to the increase in the amount of readout hardware. While such a configuration would present a much more challenging input data rate and have more significant dead time losses, the total amount of data collected should be similar to conventional radiotherapy, given the much shorter duration of FLASH irradiation to achieve doses on the same order of magnitude as a conventional fraction \autocite{matuszak_flash_2022}.
    
    \section{Conclusion}
    \label{sec:rate_conclusion}
    
    The prototype fIVI Range Monitoring System is the first device using large-area high-resolution silicon sensors which fits the rate and dynamic range requirements of clinical radiotherapy. This system demonstrates the ability to handle event rates up to \qty{1.3}{\mega\hertz} in low-energy testing, with less than \qty{0.004}{\percent} of events being rejected due to pileup or dead time. Reconstruction of the primary beamspot on the scattering target indicates spatial resolution superior to clinical requirements, as well as high tracking efficiency in both low-energy testing and at clinically-relevant beam energies, demonstrating the suitability of this detection system for the energy deposits associated with \cion{} beams on the order of \qty{200}{\mega\eV\per\nucleon}. The system also shows promise in the ability to monitor ultra-high dose rate FLASH irradiation, although this capability has not yet been demonstrated experimentally. Further measurements are planned to investigate the performance of the current prototype under FLASH conditions, with the possibility of a hardware upgrade to unlock higher data rates if required. Clinical implementation of the fIVI Range Monitoring System for radiotherapy is expected to facilitate more consistent dose delivery, allowing for a reduction in dose to healthy tissue while maintaining the same rates of tumour control as current treatments.
    
    \printbibliography[title=References]
    
\end{document}